\documentclass[final, a4paper]{IEEEtran}
\usepackage{amsmath}
\usepackage{theorem}
\usepackage{amssymb}
\usepackage{times}
\usepackage{color}
\usepackage{epsfig, epsf}
\usepackage{cite,url}

\newcommand{\ls}[1]
   {\dimen0=\fontdimen6\the\font
    \lineskip=#1\dimen0
    \advance\lineskip.5\fontdimen5\the\font
    \advance\lineskip-\dimen0
    \lineskiplimit=.9\lineskip
    \baselineskip=\lineskip
    \advance\baselineskip\dimen0
    \normallineskip\lineskip
    \normallineskiplimit\lineskiplimit
    \normalbaselineskip\baselineskip
    \ignorespaces
   }

\newcommand{\insertfig}[4]{
\begin{figure}[tbh]
\centerline{\includegraphics[width=#1\columnwidth]{#2.eps}}
\vspace{-0.3cm}
\caption{#3}\label{#4}\end{figure}}

\DeclareMathAlphabet{\mathsfbf}{OT1}{cmss}{sbc}{n}

\newcommand{\example}[2]{
\begin{center}
\parbox{0.95\columnwidth}{
\rule{0.95\columnwidth}{0.5mm}\\
\noindent {\bf Example~#1:}
#2\\
\rule{0.95\columnwidth}{0.5mm}
}
\end{center}
}

\newtheorem{lem}{Lemma}[section]

\newcommand{\EE}{\mathop{\mathbb{E}}\limits} 
\newcommand{\RR}{\mathbb{R}} 
\newcommand{\Herm}{^\dagger} 

\newcommand{\ee}{{\rm e}}
\newcommand{\jj}{{\rm j}}  
\newcommand{\dd}{{\rm\,d}} 

\newcommand{\erf}{{\rm erf}} 

\newcommand{\av}{{\bf a}}

\newcommand{\nv}{{\bf n}}

\newcommand{\pv}{{\bf p}}

\newcommand{\sv}{{\bf s}}

\newcommand{\vv}{{\bf v}}
\newcommand{\xv}{{\bf x}}

\newcommand{\zerov}{{\bf 0}}

\newcommand{\Am}{{\bf A}}
\newcommand{\Bm}{{\bf B}}
\newcommand{\Cm}{{\bf C}}

\newcommand{\Gm}{{\bf G}}

\newcommand{\Id}{{\bf I}}

\newcommand{\Rm}{{\bf R}}

\newcommand{\Wm}{{\bf W}}

\newcommand{\Xm}{{\bf X}}
\newcommand{\Ym}{{\bf Y}}
\newcommand{\Zm}{{\bf Z}}

\def\xvh{{\hat{\xv}}}

\newcommand{\deltav}{\boldsymbol{\delta}}

\newcommand{\Deltam}{\hbox{\boldmath$\Delta$}}

\newcommand{\Psim}{\hbox{\boldmath$\Psi$}}

\newcommand{\diag}{{\hbox{diag}}}

\def\trace{\mathsf{Tr}}

\newcommand{\eqdef}{\ensuremath{\stackrel{\mbox{\upshape\tiny $\Delta$}}{=}}}
\def\ben{\begin{enumerate}}
\def\beq{\begin{equation}}
\def\beqa{\begin{eqnarray}}
\def\bit{\begin{itemize}}
\def\een{\end{enumerate}}
\def\eeq{\end{equation}}
\def\eeqa{\end{eqnarray}}
\def\eit{\end{itemize}}

\def\non{\nonumber\\}

\def\MSEav{{\rm MSE}_{\rm av}}
\def\MSEinf{{\rm MSE}_{\infty}}
\def\limbeta{\lim_{\substack{M,r\rightarrow +\infty \\\beta}}}
\def\limbetaomega{\lim_{\substack{M,r\rightarrow +\infty \\ \sigma_\delta\rightarrow 0
                         \\\beta, \omega}}}

\title{Performance of Linear Field Reconstruction\\ Techniques
with Noise and Uncertain Sensor Locations
}
\author{Alessandro Nordio$^\star$, Carla-Fabiana Chiasserini, Emanuele Viterbo\\
Dipartimento di Elettronica, Politecnico di Torino\\
C. Duca degli Abruzzi 24, I-10129 Torino, Italy\\
Phone: +39 0115644183, Fax: +39 0115644099
E-mail: \tt{\{alessandro.nordio,carla.chiasserini,emanuele.viterbo\}@polito.it} \\
{\rm $^\star$ Corresponding author}}

\begin{document}

\maketitle

\begin{abstract}
We consider a wireless sensor network, sampling a 
bandlimited field, described by a limited number of harmonics.
Sensor nodes are irregularly deployed over the area of
interest or subject to random motion; in addition sensors
measurements are affected by noise. Our goal is to obtain a high
quality reconstruction of the field, with the mean square error
(MSE) of the estimate as performance metric. In particular, we
analytically derive the performance of several
reconstruction/estimation techniques based on linear filtering. For
each technique, we obtain the MSE, as well as its
asymptotic expression in the case where the field
number of harmonics and the number of sensors grow
to infinity, while their ratio is kept constant.
Through numerical simulations, we show the validity of the
asymptotic analysis, even for a small number of sensors.
We provide some novel guidelines for the design of sensor networks when many parameters,
such as field bandwidth, number of sensors, reconstruction quality,
sensor motion characteristics, and noise level of the measures, have
to be traded off.
\end{abstract}

{\bf EDICS:}  DSP-RECO Signal reconstruction, DSP-SAMP Sampling,
SEN-APPL Applications of sensor networks,
SEN-FUSE Data fusion from multiple sensors,
SPC-PERF Performance analysis and bounds

\newpage
\section{Introduction}
Wireless sensor networks are often used for applications like environmental and traffic control,
habitat monitoring, or weather forecasts~\cite{Akyildiz02}, which
require to sample a physical phenomenon over an area of interest
(the sensor field).
In this paper, we consider a set of
sensors communicating with a sink node,
through either single- or multi-hop communications.
Each sensor locally samples the physical field, while the sink collecting all samples
is in charge of reconstructing the signal of interest.

We assume that initially sensors are either located at pre-defined positions, or, if randomly deployed over the
network area, their location
can be estimated at the sink node (see~\cite{Hightower01,Hu04,Moore04} for
a description of node location methods in sensor networks).
We do not deal with spatio-temporal correlation, but
consider a fixed time instant and focus on
the spatial sampling and reconstruction of the sensor field.
We note that, in general, sensors provide an irregular sampling of the observed phenomenon.
This may be due to various reasons: random deployment of the nodes,
environment characteristics that bias the network deployment,
sensors entering a sleep mode, inaccuracy in sensor positioning, or
nodes movement~\cite{Ganesan03}. In all these cases the sink has to
reconstruct the field from a collection of samples that are {\em
irregularly} spaced, different from the classical equally (or {\em regularly}) spaced sampling.

The problem of signal reconstruction from irregular samples has been
widely addressed in signal processing, where several efficient and
fast algorithms have been proposed to numerically reconstruct or
approximate a signal \cite{Feichtinger95,Grochenig99}. The problem
we address in this work, however, is different; the questions we
pose are:
\begin{quote}
{\em (i) How do noisy measures and inaccurate knowledge of
the sensor positions affect the quality of the reconstructed signal?}\\
{\em (ii) How can we trade off system parameters like measurement
noise, field bandwidth, signal reconstruction quality and number of
sensors?}
\end{quote}

To answer these questions we analyze two different models of the monitoring system that account for
the quality of the measurements performed by the sensors and differ in
the accuracy with which the sensor positions are known at the sink node.
More specifically, the model denoted as {\bf Model A} refers to the case
where sensors are fixed, the sink has perfect knowledge of the sensor positions,
but the sensor measurements are affected by error. In the second model, named {\bf Model B},
besides noisy measurements, we consider that the sensors position varies around an average
value, and only the average location of the nodes is known at the sink.
Examples where this model applies are observation systems using surface
buoys~\cite{AOSN}, underwater robots located at different depths~\cite{Bokser04,Cayirci06},
dropsondes or low-cost unmanned platforms, as in \cite{Majumdar06}.

For each of these models, we use as field
reconstruction techniques some linear filters that are commonly
employed in signal detection and estimation, and we evaluate the
mean square error of the resulting estimate.

We find that a key parameter for the network performance is the
ratio $\beta$ of  the field number of harmonics to the number of sampling
sensors. In particular, there exists a value of this ratio,
beyond which the performance of all considered reconstruction
strategies degrade significantly, even for low values of noise level
and limited uncertainty on the sensor positions. To obtain an
acceptable reconstruction quality when $\beta$ is large (i.e., the
number of available sensors is limited compared to the field
bandwidth), reconstruction techniques that exploit some knowledge of
the measurement noise and of the jitter in the sensors position must be employed.

We also carry out an asymptotic analysis of the system as the field
number of harmonics and the number of sensors grow to infinity, while their
ratio $\beta$ is kept constant, and we show that this is an
effective tool to study the system performance, even when the number
of sensors is small. Finally, we find a lower bound to the mean
square error that can be achieved by any of the considered
techniques, both under Model A and Model B.

The remainder of the paper is organized as follows.
In Section~\ref{sec:system_model} we present our assumptions and
the system models under study. Section \ref{sec:related-work}
highlights our contribution with respect to previous work.
Section~\ref{sec:defs_perf} introduces the performance metrics and
provides some mathematical tools necessary for our study. Model A
and B are analyzed in Sections~\ref{sec:model_A}
and~\ref{sec:model_C}, respectively.
Finally, in Sections \ref{sec:summary} and \ref{sec:conclusions} we summarize our main results and
draw some conclusions.

\section{Assumptions and System Models}
\label{sec:system_model}

Let us consider a one-dimensional bandlimited field
$s(x)$ represented by $2M+1$ harmonics as
\begin{equation}
s(x) = \frac{1}{\sqrt{2M+1}}\sum_{k=-M}^M a_k\ee^{\jj 2 \pi k x}
\end{equation}
The field is observed within one period interval $[0,1)$ and sampled by $r$ sensors
placed at positions%
\footnote{Column vectors are denoted by bold lowercase
letters, matrices are denoted by bold upper case letters. The
$(k,q)$ entry of the matrix $\Xm$ is denoted by $(\Xm)_{kq}$. The
$n\times n$ identity matrix is denoted by $\Id_n$, the generic
identity matrix is denoted by $\Id$, and the conjugate transpose
operator is denoted by $(\cdot)\Herm$}
$\xv=[x_1,\ldots,x_r]^{\rm T}$, $x_q\in [0,1)$, $q=1,\ldots,r$
which are in general not equally spaced.
The signal samples are denoted by the column vector $\sv=[s(x_1), \ldots, s(x_r)]^{\rm T}$.
The field discrete spectrum is given by the $2M+1$ complex
vector $\av=[a_{-M}, \ldots, a_0, \ldots, a_M]^T$. The complex numbers $a_k$ represent amplitudes and phases
of the harmonics in $s(x)$. We can think of $M$ as the approximate one-sided bandwidth of the field.

We assume that the entries of $\xv$ are
i.i.d. uniformly distributed random variables in $[0,1)$. The
extension to a multi-dimensional field can be easily obtained, as
discussed later in this section.

We define $\beta$ as the ratio of the number of harmonics
which describe the field to the number of sensors, i.e., $\beta=(2M+1)/r$. This is an
important parameter in our analysis.
Note that the number of sensors $r$ also corresponds to the sampling rate;
thus, the number $\beta$ is the ratio of twice the field bandwidth
to the sampling rate (frequency).
In particular, in regular sampling theory, exact reconstruction is achieved for $\beta \in [0,1)$ and,
if a Nyquist regular sampling interval were used, we would have: $\beta=1$.

We consider $M$ to be known, and the random vector $\av$ to have zero mean and covariance matrix
$\EE[\av\av\Herm]=\sigma^2_a\Id_{2M+1}$, where $\sigma^2_a$ corresponds to the field
average power spectral density.

The value $\sv$ of the field at positions $\xv$ depends on the spectrum $\av$
through the expression
\begin{equation}
\sv =\Gm_\xv\Herm\av
\label{eq:s}
\end{equation}
where $\Gm_\xv$ is the $(2M+1)\times r$
{\em generalized Fourier matrix} defined as:
\begin{equation}
(\Gm_\xv)_{kq}=\frac{1}{\sqrt{2M+1}}\ee^{-\jj 2\pi  k x_q}~~~
\begin{array}{l} k= -M,\ldots, M \\
q=1,\ldots,r\end{array}
\label{eq:F}
\end{equation}
The dependence of the matrix $\Gm_\xv$ on the position vector $\xv$
is clearly indicated by its subscript.
When the samples are equally spaced in the interval $[0,1)$,
the matrix $\sqrt{\beta}\Gm_{\xv}$ is a unitary matrix
(i.e., $\beta\,\Gm_{\xv}\Gm^{\dagger}_{\xv}=\Id_{2M+1}$).
The above system model refers to a uni-dimensional
field where sensor positions are determined by a scalar variable.
However, the extension to the multi-dimensional case can still be
easily obtained since the relation between field spectrum and
samples in a band-limited multi-dimensional field can be expressed
in a matrix form similar to (\ref{eq:s}), where only the structure
of the matrix $\Gm_\xv$ differs.

Finally, we assume that sensor field measures are sent to a processing unit,
the so-called {\em sink} node, whose task is to provide an estimate
of the sensed field.
Since we focus on the reconstruction of the physical field,
we consider that sensor transmissions always reach successfully the sink node%
\footnote{Note that this is a fair assumption since, when ARQ or FEC techniques are used, the information either
is correctly retrieved at the sink or it is lost. The latter case corresponds to reduced value of $r$
}.

By relying on the assumptions discussed above, we study the following two systems.
\begin{itemize}
\item  {\bf Model A: Fixed sensors, perfect knowledge of the sensor positions, noisy measures}\\
In this model, sensors have a fixed position, given by the vector
$\xv$ and known at the sink node, but each sensor provides a measure
of the field affected by additive noise with zero mean and variance
$\sigma^2_n$~\cite{Vuran04}.
The additive noise approximates the errors affecting the measurement procedure~\cite{Ergen06}.

The measures vector can therefore be written as:
\begin{equation}
\pv = \sv+\nv = \Gm_\xv\Herm \av + \nv
\label{eq:p_fixed}
\end{equation}
where $\sv$ is the true field and the zero mean noise vector is denoted by
$\nv$, with covariance matrix $\EE[\nv\nv\Herm]=\sigma^2_n\Id_r$.

\item {\bf Model B: Sensors with jittered positions and noisy measures}\\
In this case each sensor moves around an average position
$\hat{x}_q$ ($q=1,\ldots,r$), i.e., the sensor positions are given
by: $\xv = \xvh + \deltav$, where: $\EE[\xv]=\xvh$ and $\deltav$ is
the displacement of the sensors with respect to their average
location $\xvh$.  Note that our problem differs from
the well known problem of jittered sampling (see e.g.,
\cite{LiuStanley}), since we deal with irregular sample locations.
The displacements $\delta_q$, $q=1,\ldots,r$, are modeled as
independent zero mean Gaussian random variables with variance
$\sigma^2_\delta$ and $\EE[\deltav\deltav\Herm]
=\sigma^2_\delta\,\Id_r$.  For convenience and
neglecting the edge effects, we consider $\mod(x_q,1)$ so that $x_q$
falls in the observation interval $[0,1)$. The vector $\pv$ of
measures is still given again by~(\ref{eq:p_fixed}). Also,  noise,
displacement, and field spectrum are assumed to be 
uncorrelated, hence $\EE[\nv\deltav\Herm] = \EE[\nv\av\Herm]=
\EE[\av\deltav\Herm] = \zerov$, and the sink has perfect knowledge
of $\xvh$.
\end{itemize}

\section{Our contribution with respect to previous work}
\label{sec:related-work}

Given a network where sensors can enter a low-power operational
state (i.e., a sleep mode), the work in \cite{Perillo04}  presents an algorithm to determine which
sensor subsets should be selected to acquire data from an
area of interest and which nodes should remain inactive to
save energy.
A similar problem is addressed in \cite{Willett04}, where an adaptive sampling is
described, which
allows the central data-collector to vary the number of active
sensors, i.e., samples, according to the desired resolution level.
The optimal sensor density that minimizes the network energy consumption,
subject to constraints on the quality of the reconstructed signal and
network lifetime, is studied in \cite{Maleki05}.
Note that in our work we consider an irregular topology, which may be
caused by nodes moving into a sleep state; however we do not directly
address energy efficiency or scheduling of the node sleep/activity
periods.

In \cite{Kumar03}, the authors consider a uni-dimensional field, uniformly sampled at the
Nyquist frequency by low-precision sensors. The impact on the
field reconstruction accuracy of quantization errors
and node density is evaluated.
The effect of random error sources affecting the ADC, besides quantization,
is investigated in \cite{Ergen06}.
In our work we consider an additive noise that models errors due to the measurement
procedure as well as errors due to the ADC, but we do not specifically focus
on the latter issue.

The impact of medium access control (MAC) protocols on
the reconstruction of a signal field is investigated in \cite{Dong04}.
Both deterministic and random MAC schemes are considered,
and performance are derived as the number of received packets and
the experienced SNR vary.

Related to our work is also the literature on
spectral analysis~\cite{Maravic, Stoica}, which
deals with the problem of recovering the amplitude of sine waves
immersed in noise.
Note, however, that techniques such as MUSIC do not estimate phases; thus,
we do not compare with such techniques since our linear filtering reconstruction yields the
estimate of both amplitudes and phases.

The field reconstruction at the sink node with spatial and
temporal correlation among sensor measures is studied in
\cite{CristescuVetterli,Poor,Vuran04,Rachlin1}.
In particular, in~\cite{Rachlin1} the observed field is a discrete vector of
target positions and sensor observations are dependent. By modeling
the sensor network as a channel encoder and exploiting some concepts
from coding theory, the network capacity,
defined as the maximum ratio of target positions to number of sensors,
is studied as a function of noise, sensing function and sensor connections.
The paper by Dong and Tong~\cite{DongTong} focuses on signal reconstruction from possibly random
samples, as we do. However, two major issues make our work significantly
different from~\cite{DongTong}. Dong and Tong indeed assume that the exact
sensors locations are  known and that the central controller
always receives a sufficiently large number of samples. These assumptions allow
an interpolation method, which is used in  Dong and Tong's work,
to provide good performances. In our case, instead, even
in the asymptotic analysis, the ratio of the number of harmonics
to the number of samples is kept constant and, hence, interpolation
may be highly inefficient as we will show in the following.

The problem of reconstructing a band-limited signal from an
irregular set of samples at unknown locations is addressed
in~\cite{Marziliano00}. There the signal is oversampled by
irregularly spaced sensors; sensor positions are unknown but always
equal to an integer multiple of the sampling interval. Different
solution methods are proposed, and the conditions for which there
exist multiple solutions or a unique solution are discussed.
Differently from~\cite{Marziliano00}, we assume that the sink can
either acquire or estimate the sensor locations and that sensors are
randomly deployed over a finite interval.

Finally, in our previous work \cite{NordioChiasseriniViterbo} some conditions
on the irregular topology of the sensor network are identified, which allow for
a successful signal reconstruction, both under deterministic and random
node deployment. In particular, in \cite{NordioChiasseriniViterbo} the
spectrum estimate $\hat{\av}$,
computed by the sink, is obtained by applying to $\sv$ the
Moore-Penrose pseudo-inverse of the matrix $\Gm_\xv$, i.e.,
$\hat{\av} = \left(\Gm_\xv\Gm_\xv^\dagger \right)^{-1}\Gm_\xv\sv$.
The system model adopted in \cite{NordioChiasseriniViterbo} is ideal in the sense that
the reconstruction algorithm has perfect knowledge of the vector $\xv$ and
neglects noisy measures: the failure in reconstruction (i.e. $\hat{\av}\neq \av$)
is only due to the bad conditioning of the matrix $\Gm_\xv\Gm_\xv\Herm$ in relation to the
finite machine precision.
In this work, instead, we propose to apply linear filters to the field reconstruction
and consider the following causes of quality degradation: {\em (i)}
noisy measures, and {\em (ii)} uncertainty at the sink on the sensors position.

\section{Preliminaries}
\label{sec:defs_perf}

Here we describe the techniques we use for field reconstruction,
and define the performance metrics employed for assessing
the effectiveness of these techniques on the quality of the reconstructed field.
Finally, we provide some mathematical tools necessary for the analysis of the
models under study.

\subsection{Reconstruction techniques}
Several reconstruction techniques have been proposed in the literature,
which  amount to the solution of a linear system
(see \cite{Feichtinger95,Grochenig99} and the references cited therein). A
widely used technique consists in processing the measures $\pv$ by means of
a linear filter,
$\Bm$, which is an $r \times (2M+1)$ matrix and is a function of the system parameters known
at the sink. In this case, the estimate of the field spectrum is given by:
\begin{equation}
\hat{\av} = \Bm\Herm\pv
\label{eq:hat_a}
\end{equation}
The system model in~(\ref{eq:p_fixed}) is similar
the one employed in multiuser communications~\cite{Verdu_book} or multiple antennas
communications \cite{TulinoVerdu,BigTarTul}.
In those cases $\pv$ is the received signal, the matrix $\Gm_\xv$ plays the role of
spreading matrix or channel matrix, $\av$ is the transmitted signal and $\nv$ is
the channel noise.
By relying on the results obtained in those fields, for each system
model we propose and compare some reconstruction techniques
characterized by different matrices $\Bm$: the {\em matched filter
(MF)}, the {\em zero forcing (ZF)} filter and some {\em linear filters
minimizing the mean square error} (LMMSE)~\cite{Verdu_book}. In the
field of multiuser detection, the MF simply correlates the received
signal with the desired user's time reversed spreading waveform,
thus it does not take into account any other users in the system or
channel dynamics. The ZF filter counteracts multiuser interference
but it ignores the presence of channel noise. The LMMSE solution
minimizes the squared error between the received and transmitted
signals, thus accounting also for the channel noise; it becomes the
zero-forcing solution when no noise is present. Note that the
advantage of the MF with respect to the ZF and LMMSE filters is that
no matrix inversion is needed; while, between the ZF and the LMMSE
filter, clearly the best performance in terms of minimum square
error is given by the LMMSE, however the advantage of the ZF filter
is that it does not require any knowledge of the noise component
(see \cite{Verdu_book} for further details).

\subsection{Performance metrics}
\label{sec:MSE}
Given the spectrum estimate~(\ref{eq:hat_a}), the field can be reconstructed
as:
\[ \hat{s}(x) = \frac{1}{\sqrt{2M+1}}\sum_{k=-M}^M \hat{a}_k\ee^{\jj 2 \pi k x} \]
As a measure of the quality of the reconstruction, we consider
the {\em mean square error} (MSE) of the estimate of $s(x)$, which is given by:
\[ {\rm MSE} = \EE\left[\int_0^1|s(x)-\hat{s}(x)|^2 \dd x \right] \]
We observe that computing MSE as above is equivalent to computing $\EE[\|\av - \hat{\av} \|^2]$. 
Indeed, we have:
\begin{eqnarray}
&&\int_0^1|s(x)-\hat{s}(x)|^2 \dd x \non
&=& \frac{1}{2M+1}\int_0^1\left|\sum_{k=-M}^M\ee^{\jj 2 \pi k x}(a_k-\hat{a}_k)   \right|^2 \dd x \non
&=& \frac{1}{2M+1}\sum_{\substack{k=-M\\ h=-M}}^M\int_0^1\ee^{\jj 2 \pi (k-h) x} \dd x\, (a_k-\hat{a}_k)(a_h-\hat{a}_h)^* \non
&=& \frac{1}{2M+1}\sum_{k=-M}^M  |a_k-\hat{a}_k|^2 =\frac{1}{2M+1}\|\av-\hat{\av}\|^2 \nonumber
\end{eqnarray}
Therefore, in the following, for a given vector of sensor positions $\xv$, we consider the MSE defined as:
\begin{equation}
{\rm MSE}_\xv = \frac{\EE\left[\|\hat{\av} -\av\|^2\right]}{2M+1} = \frac{\sigma^2_a}{2M+1}\trace\{\Psim_\xv\}
\label{eq:MSE}
\end{equation}

where
\begin{equation}
\Psim_\xv \eqdef \frac{1}{\sigma^2_a}\EE\left[(\hat{\av} -\av)(\hat{\av} -\av)\Herm\right]
\label{eq:Psi}
\end{equation}
is a $(2M+1)\times (2M+1)$ matrix,
the operator $\EE[\cdot]$ averages with respect to all random variables of the model, and
$\trace\{\cdot\}$ is the trace operator.
Also, in (\ref{eq:MSE}) we exploited the fact that, for any vector $\vv$,
we have: $\EE\left[\|\vv\|^2\right]= \EE\left[\trace\{\vv\vv\Herm\}\right]
= \trace\left\{\EE\left[\vv\vv\Herm\right]\right\}$.

Next, we consider the vector $\xv$ to be random.
In this case a more appropriate performance metric is the
{\em average} MSE, normalized to $\sigma^2_a$, i.e.,
\[ \MSEav = \frac{\EE_{\xv}[{\rm MSE_\xv}]}{\sigma^2_a} \]
where MSE$_\xv$ is as in (\ref{eq:MSE}) and $\EE_\xv[\cdot]$ averages over the realizations of $\xv$.

When the parameters
$M$ and $r$ grow to infinity while the ratio $\beta=(2M+1)/r$ is kept constant, we
define the {\em asymptotic average} MSE as:
\begin{equation}
\MSEinf = \lim_{\substack{M,r\rightarrow
+\infty \\\frac{2M+1}{r}=\beta}} \MSEav
\label{eq:MSEinf}
\end{equation}
Our results will show later that $\MSEinf$ gives a very good approximation of MSE$_{av}$ already for small
values of $M$. This is a common feature of asymptotic analysis based on random matrices~\cite{TulinoVerdu}.

\subsection{Some mathematical tools}

\subsubsection{The functional $\phi$}
\label{sec:free|probability}

Let us first consider an $n\times n$ Hermitian random matrix $\Xm$ and the functional:
\[ \phi(\Xm) \eqdef \lim_{n\rightarrow +\infty} \frac{1}{n}\EE[\trace\{\Xm\}] \]
Using~(\ref{eq:Psi}) and~(\ref{eq:MSEinf}), the asymptotic MSE can be written as:
\begin{equation}
{\MSEinf} = \phi(\Psim_\xv)
\label{eq:MSEinf-Psi}
\end{equation}
In our analysis we use  the following results on the functional
$\phi(\cdot)$. First, we notice that: $\phi(\Id)=1$. Secondly, we
can prove that, if $g(x)$ is an analytic function defined in $x
>0$ then%
\footnote{
Note the small abuse of notation
when using $g(\cdot)$ for both scalar and matrix argument
}
\begin{equation}
\phi(g(\Xm)) = \EE\left[g(\xi)\right]
\label{eq:g(X)}
\end{equation}
where $\xi$ is a random variable with the asymptotic eigenvalue distribution of $\Xm$.
The proof is given in Appendix~\ref{app:proof3}.

\subsubsection{A simple expression for $\Gm_{\xvh+\boldsymbol{\delta}}$}
As will be clear in Section~\ref{sec:model_C},
in the analysis of Model B many parameters are functions of
the matrix $\Gm_\xv$, where $\xv=\xvh+\deltav$.
It is thus useful to derive an expression of $\Gm_\xv$ as a
function of $\Gm_\xvh$, in order to separate the random part $\deltav$ of $\xv$
from the constant part $\xvh$. From~(\ref{eq:F}), the $(k,q)$ entry of $\Gm_{\xv}$
is defined as:
\[ (\Gm_{\xv})_{kq} = \frac{1}{\sqrt{2M+1}}\ee^{-\jj 2\pi k x_q} =
     \frac{1}{\sqrt{2M+1}}\ee^{-\jj 2\pi k \hat{x}_q} \,\ee^{-\jj 2\pi k \delta_q} \]
A useful expression of $\Gm_{\xvh}$ in terms of $\Gm_\xv$ is given below.

\begin{lem}
\label{lemma:1}
For any vector $\xv$ of size $r$, let the $(k,q)$ entry of the matrix
$\Gm_\xv$ be
\[ (\Gm_\xv)_{kq} = \frac{1}{\sqrt{2M+1}}\ee^{-\jj 2\pi k x_q} \]
for $k=-M,\ldots,M$, and $q=1,\ldots,r$. Let the size $r$ column vectors
$\xv$, $\xvh$, and $\deltav$ be such that $\xv = \xvh +\deltav$, then
\begin{equation}
\Gm_{\xv} = \sum_{n=0}^\infty \frac{1}{n!}\Wm^n \Gm_\xvh \Deltam^n
\label{eq:theorem_equation}
\end{equation}
where $\Deltam = {\rm diag}(\deltav)$ is an $r\times r$ diagonal matrix, and
$\Wm$ is a $(2M+1)\times (2M+1)$ diagonal matrix with $(\Wm)_{kk} = -\jj 2\pi k$.
\end{lem}
\medskip
\noindent {\it Proof:} The proof is given in Appendix~\ref{app:proof}.

\section{Analysis of Model A}
\label{sec:model_A}

Here we consider the case where sensor positions are fixed and known
at the sink but the field estimates are degraded by noisy measures.
We analyze three different linear filters: the matched filter, the
zero forcing filter and the minimum mean square error filter
\cite{Verdu_book}.
In all cases, for any fixed $\xv$, the filter matrix $\Bm$ is deterministic.
Thus, using (\ref{eq:p_fixed}),~(\ref{eq:hat_a}), and~(\ref{eq:Psi}) we obtain:
\begin{eqnarray}
\Psim_\xv
&=& \frac{1}{\sigma^2_a}\EE_{\av,\nv}\left[\|\hat{\av}-\av\|^2\right] \non
&=& \frac{1}{\sigma^2_a}\EE_{\av,\nv}\left[\|\Bm\Herm(\Gm_\xv\Herm\av+\nv)-\av\|^2\right] \non
&=& (\Bm\Herm\Gm_\xv\Herm-\Id)(\Gm_\xv\Bm-\Id) +\alpha\Bm\Herm\Bm
\label{eq:MSE_A}
\end{eqnarray}
where
\[ {\rm SNR}_m = \frac{1}{\alpha}=\frac{\sigma^2_a}{\sigma^2_n}\]
is the {\em signal-to-noise ratio on the measure}.
The MSE expression specialized to the different filters is given below.


\subsection{Matched filter}
As a first solution, we choose $\Bm$ as the filter matched to
$\Gm_\xv$. The MF is optimal  when the collected samples are equally
spaced,  that is when the rows of $\sqrt{\beta}\,\Gm_{\xv}$ are orthonormal
vectors and $\sqrt{\beta}\,\Gm_{\xv}$ is a unitary matrix (i.e.,
$\beta\,\Gm_{\xv}\Gm_{\xv}^\dagger=\Id_{2M+1}$). Thus, we choose:
\begin{equation}
\Bm\Herm = \beta\Gm_{\xv}
\label{eq:A_B_MF}
\end{equation}
Recall that $\Gm_{\xv}$ depends on the position vector $\xv$ that,
under Model A, coincides with the actual sensor positions.
Indeed, in the absence of noise and for equally spaced sensors, we have
the spectrum estimates perfectly match $\av$, i.e.,
\[ \hat{\av} = \Bm\Herm\pv = \beta\,\Gm_\xv\Herm\Gm_\xv\av = \av \]

By replacing (\ref{eq:A_B_MF}) in (\ref{eq:MSE_A}), we obtain the following
expression for $\Psim_\xv$:
\begin{equation}
\Psim_\xv = \beta^2\,\Rm_\xv^{2}+\Id+(\alpha\beta-2)\beta\Rm_\xv
\label{eq:MSE_A_MF}
\end{equation}
where $\Rm_\xv=\Gm_\xv\Gm_\xv\Herm$.
From the definition in~(\ref{eq:MSEinf-Psi}), the asymptotic MSE,
averaged over the random vector $\xv$, is given by:
\begin{eqnarray}
\MSEinf
&=& \phi(\Psim_\xv)\non
&=& \beta^2\phi(\Rm_\xv^{2}) +\phi(\Id) +(\alpha\beta-2)\beta\phi(\Rm_\xv) \nonumber
\end{eqnarray}
Notice that the second term on the right hand side reduces to $1$ since $\phi(\Id)=1$.
Applying~(\ref{eq:g(X)}),  first with $g(x)=x^2$ and then with $g(x)=x$, we obtain:
\begin{equation}
\MSEinf = \EE[\lambda^2] + 1 +(\alpha\beta-2)\,\EE[\lambda]
\label{eq:MSEinf_A_MF}
\end{equation}
where $\lambda>0$ is the random variable with probability density function (pdf)
$f_{\lambda,\beta}(x)$, distributed as the asymptotic eigenvalues of $\beta\Rm_\xv$.
 In~\cite{NordioChiasseriniViterbo} it is shown that, for any positive integer $p$,
$\EE[\lambda^p]$ is a polynomial in $\beta$ of degree $p-1$.
In particular $\EE[\lambda]=1$ and $\EE[\lambda^2]=1+\beta$. We therefore obtain:
\begin{equation}
\MSEinf = \beta(\alpha+1)
\label{eq:MSEinf_A_MF2}
\end{equation}


\subsection{ZF filter}
\label{sec:A_ZF}
The expression of the ZF filter for the system in~(\ref{eq:p_fixed}) is:
\begin{equation}
\Bm\Herm = \Rm_\xv^{-1}\Gm_\xv
\label{eq:A_B_ZF}
\end{equation}
Notice that, by its definition, the ZF filter does not exploit any information on the
noise contribution (such as $\sigma^2_n$). However,  this
reconstruction technique
takes into account the fact that the collected samples are not equally spaced and, hence,
that  $\sqrt{\beta}\,\Gm_\xv$ is not a unitary matrix.

By using (\ref{eq:A_B_ZF}) in (\ref{eq:MSE_A}), the matrix $\Psim_\xv$ becomes:
\begin{equation}
\Psim_\xv = \alpha\Rm_\xv^{-1}
\label{eq:MSE_A_ZF}
\end{equation}
Using the definition in~(\ref{eq:MSEinf-Psi}) and applying~(\ref{eq:g(X)})
with $g(x)=x^{-1}$, the asymptotic MSE,
averaged over the random vector $\xv$, can be written as:
\begin{equation}
\MSEinf = \alpha\,\phi(\Rm_\xv^{-1}) = \alpha\beta\,\EE\left[\frac{1}{\lambda}\right]
\label{eq:MSEinf_A_ZF}
\end{equation}
We can make the following observations on the behavior of the $\MSEinf$:
\begin{itemize}
\item[{\em 1)}] Since $1/\lambda$ is a convex function, then $\EE[1/\lambda] \ge 1/\EE[\lambda]$.
In~\cite{NordioChiasseriniViterbo} it is shown that $\EE[\lambda]$=1, thus it results: $\MSEinf \ge \alpha\beta$.

\item[{\em 2)}] We have:
$\MSEinf =\alpha\beta\, \EE[\lambda^{-1}]<+\infty$ only for $\beta \in [0,\beta^\star)$, with $\beta^\star \approx 0.35$.
Indeed
\[ \EE\left[\frac{1}{\lambda}\right] = \int_0^{+\infty} \frac{1}{x} f_{\lambda,\beta}(x) \dd x \]

In~\cite{NordioChiasseriniViterbo} it has been  empirically observed through Monte-Carlo simulation
that for $x \ll 1$:
\[ f_{\lambda,\beta}(x)  \propto x^{a(\beta)-1} \]
where the exponent $a(\beta)$ is a decreasing function of $\beta$ for
$\beta\in [0,1]$, and $a(\beta)=1$ for $\beta=\beta^\star$.
Given that, for any positive constant $c$, we have:
\[ \int_0^c \frac{1}{x} f_{\lambda,\beta}(x) \dd x \propto \int_0^c x^{a(\beta)-2} \dd x \]
where the integral in the right hand side (and
therefore~(\ref{eq:MSEinf_A_ZF})) does not diverge if and only if
$a(\beta)>1$, that is $\beta<\beta^{\star}$. This observation gives
us a fundamental limit to the minimum number of sensors required to
perform reliable reconstruction with the ZF filter.
\end{itemize}

\subsection{LMMSE linear filter}
\label{sec:A_LMMSE}
A more efficient solution is to employ the filter $\Bm$ that provides the
minimum MSE (LMMSE).
By assuming that the signal-to-noise ratio ${\rm SNR}_m$ is known to the sink and exploiting
this information for the filter design,
the expression of the LMMSE filter~\cite{Verdu_book} for
Model A in~(\ref{eq:p_fixed}) is given by:
\begin{equation}
\Bm\Herm =\left(\Rm_\xv +\alpha\Id\right)^{-1} \Gm_\xv
\label{eq:A_B_LMMSE}
\end{equation}
We highlight that this reconstruction technique accounts for both the fact that
the collected samples are non-uniformly spaced and the presence of the measurement noise.

Substituting (\ref{eq:A_B_LMMSE}) in~(\ref{eq:MSE_A}), we obtain:
\begin{equation}
\Psim_\xv = \alpha\left(\Rm_\xv+\alpha\Id\right)^{-1}
\label{eq:MSE_A_LMMSE}
\end{equation}
Using~(\ref{eq:g(X)}) with $g(x)=(x+\alpha\beta)^{-1}$, the asymptotic MSE is:
\begin{equation}
\MSEinf = \EE\left[\frac{\alpha\beta}{\lambda+\alpha\beta}\right]
\label{eq:MSEinf_A_LMMSE}
\end{equation}
Note that:
\begin{equation}
\EE\left[\frac{\alpha\beta}{\lambda+\alpha\beta}\right] \ge \frac{\alpha\beta}{\EE[\lambda+\alpha\beta]}
=\frac{\alpha\beta}{1+\alpha\beta}
\label{eq:MSE_bound}
\end{equation}
Also note that $\EE[\alpha\beta/(\lambda+\alpha\beta)] \le 1$, since
$\lambda\geq 0$. Given that the LMMSE filter provides the minimum
MSE, from~(\ref{eq:MSE_bound}) {\em it turns out that, for a given $\beta$},
$\alpha\beta/(1+\alpha\beta)$ {\em is a lower
bound for the performance of all linear reconstruction techniques}.

We summarize the main results of this section in Table~\ref{table:results-modelA}.

\begin{table}
\begin{center}
\caption{Results obtained under Model A \label{table:results-modelA}}
\begin{tabular}{|l||c|c|c|} \hline
              & \rule[-1.5mm]{0mm}{5mm} { MF} &  { ZF} & { LMMSE} \\ \hline
\hline
 $\Bm\Herm$   & $\Gm_\xv$&  \rule[-2mm]{0mm}{6mm} $\Rm_\xv^{-1}\Gm_\xv$ & $(\Rm_\xv +\alpha\Id)^{-1}\Gm_\xv$  \\ \hline
$\Psim_\xv$  &   $(\beta\Rm_\xv-\Id)^2+\alpha\beta^2\Rm_\xv$ &
              \rule[-2mm]{0mm}{6mm} $\alpha\Rm_\xv^{-1}$
              & $\alpha\left(\Rm_\xv+\alpha\Id\right)^{-1}$ \\ \hline
$\MSEinf$    & $\beta(\alpha+1)$ &
              \rule[-2mm]{0mm}{6mm} $\alpha\beta\,\EE\left[\frac{1}{\lambda}\right]$
              & $\EE\left[\frac{\alpha\beta}{\lambda+\alpha\beta}\right]$ \\ \hline
\end{tabular}
\end{center}
\end{table}

\subsection{Results}
In Figure~\ref{fig:MSE_model_A_0} we compare the average MSE obtained using
the MF, ZF, and LMMSE filters, when $\beta$ varies and $\alpha=1/2$ (i.e.,
SNR$_m=3$\,dB). The points labeled by ``$\MSEav$ MF'', ``$\MSEav$ ZF'' and ``$\MSEav$
LMMSE'' have been obtained generating $100$ realizations of the
measures~(\ref{eq:p_fixed}) with $M=40$, computing the estimates as
in~(\ref{eq:hat_a}) and averaging the square error $\|\av-\hat{\av}\|^2$.
These points are superimposed to the solid curves labeled by ``$\MSEinf$'',
representing the asymptotic MSE and obtained evaluating~(\ref{eq:MSEinf_A_MF}), (\ref{eq:MSEinf_A_ZF}) and
(\ref{eq:MSEinf_A_LMMSE}), respectively. Notice that computing closed
form expressions for $\EE[1/\lambda]$ in~(\ref{eq:MSEinf_A_ZF}) and
$\EE[\alpha\beta/(\lambda+\alpha\beta)]$ in~(\ref{eq:MSEinf_A_LMMSE}) is still
an open problem since a closed form expression of the distribution of $\lambda$ is unknown.
Thus, for a given $\beta$, the value of these
asymptotic expressions have been obtained pseudo-analytically, averaging over the
eigenvalues $\lambda$ obtained by several realizations of the matrix $\beta\,\Rm_\xv$, with $M=200$
which yields a very good approximation of the asymptotic case (see~\cite{NordioChiasseriniViterbo}).

We observe an excellent agreement between the asymptotic analysis
and the numerical results; this shows the validity of the asymptotic
analysis even for values of $M$ as low as $M=40$.
We also note that, for both the filters, higher values of MSE are obtained as $\beta$
increases. Finally, the LMMSE
filter provides the best performance, while the MSE of the ZF filter
shows a vertical asymptote for $\beta=\beta^\star$, in agreement
with the closed form analysis\footnote{The numerical results for the
ZF filter are highly unstable while approaching the asymptote, thus
they are shown only for $\beta \leq 0.32$}.

\insertfig{1.00}{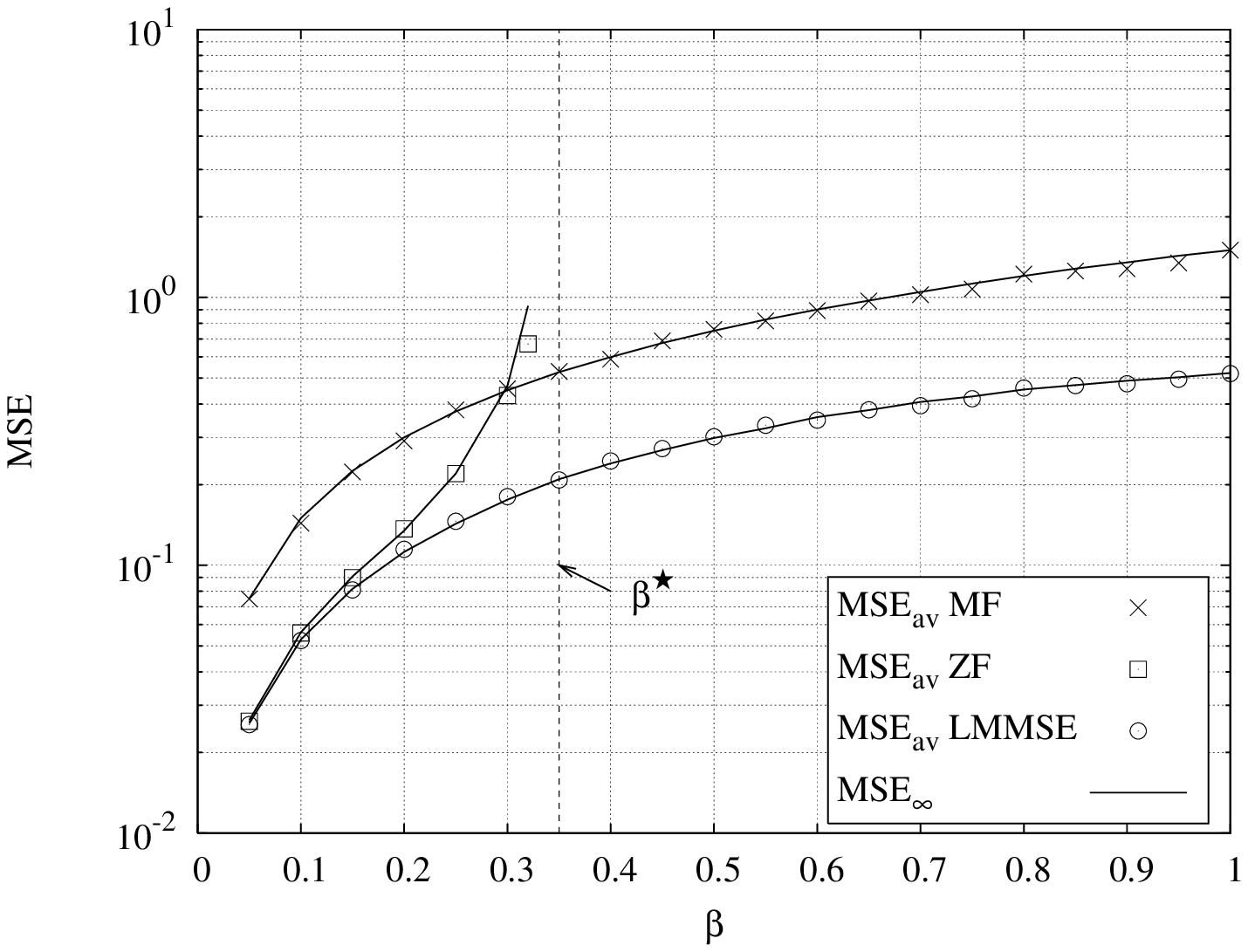}{MSE obtained through the MF, ZF, and the LMMSE filters,
plotted versus $\beta$, for $M=40$, SNR$_m=3$\,dB (i.e., $\alpha=1/2$)}{fig:MSE_model_A_0}

\insertfig{1.00}{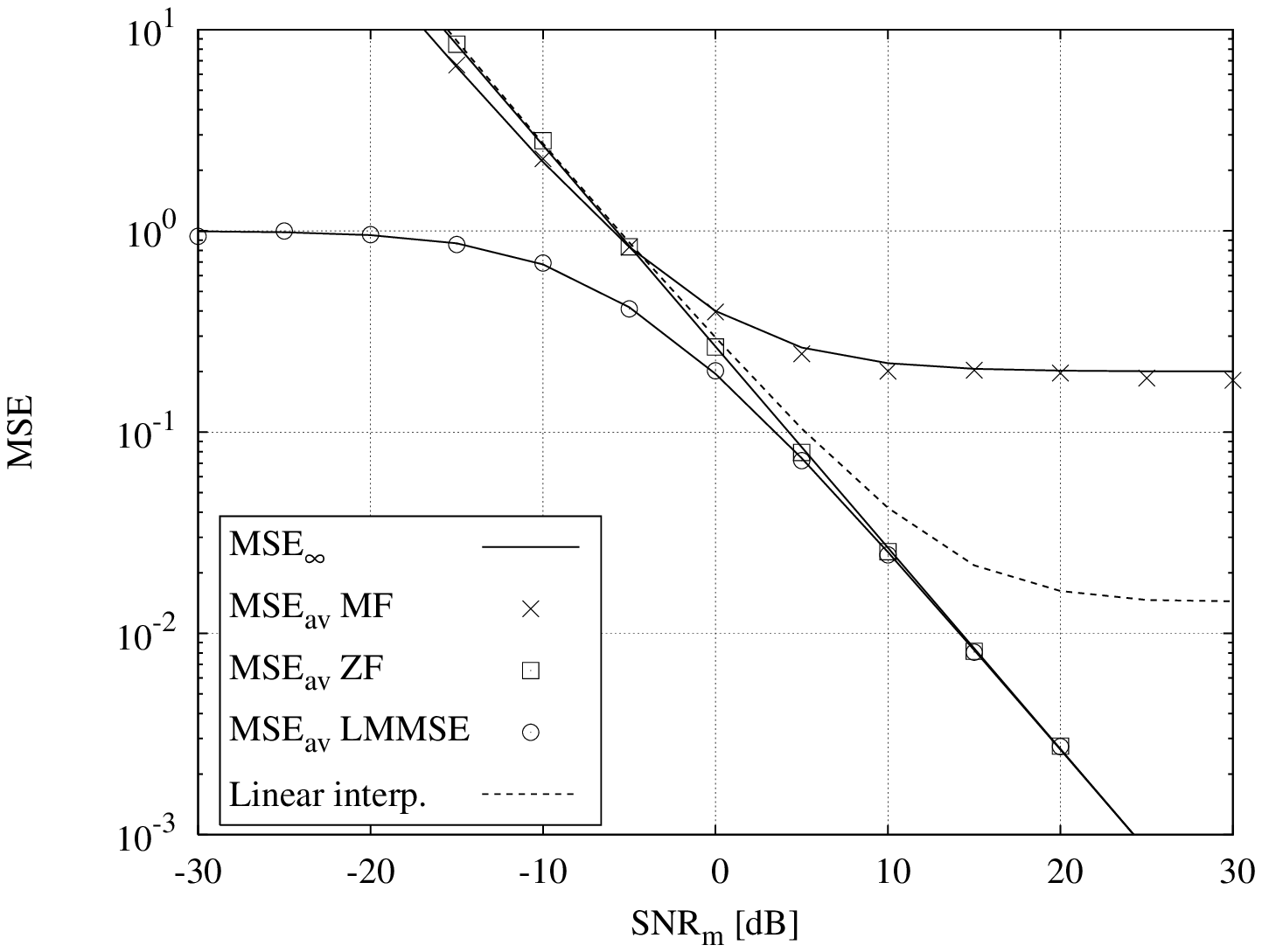}{MSE obtained with the MF, the ZF and the LMMSE filters,
plotted versus SNR$_m$, for $\beta=0.2$ and $M=10$}{fig:MSE_model_A_1}
\insertfig{1.00}{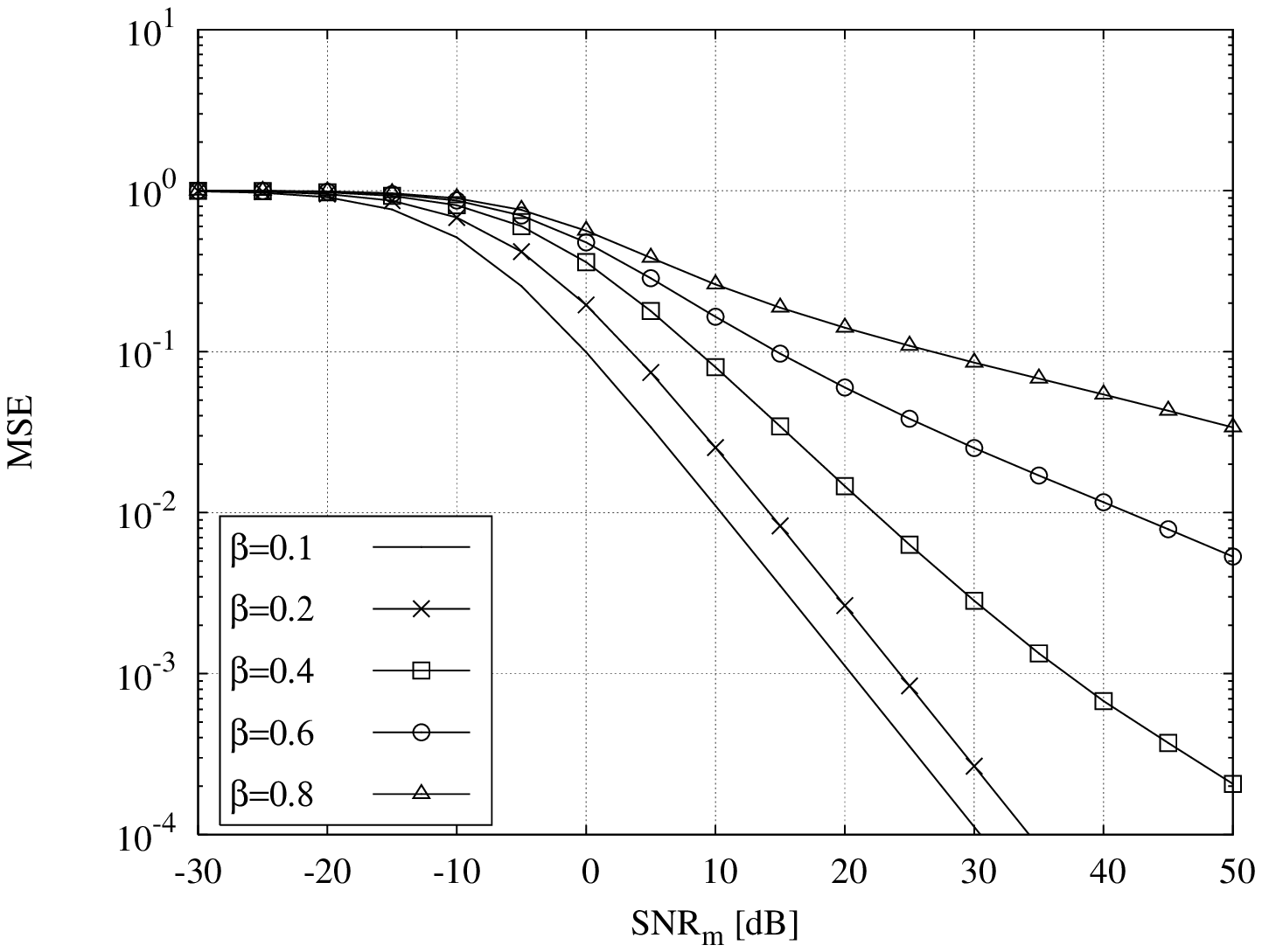}{MSE obtained through the LMMSE filter,
plotted versus SNR$_m$, for $\beta=0.1,0.2,0.4,0.6,0.8$ and $M=10$}{fig:MSE_model_A_2}

Figure~\ref{fig:MSE_model_A_1} shows the MSE versus SNR$_m$, for $\beta=0.2$.
The behavior of the asymptotic MSE is represented by the curves labeled by
``$\MSEinf$'' while the average MSE obtained through numerical analysis is denoted by
the label ``$\MSEav$''. The curves have been obtained using the same procedure as for
the results in Figure~\ref{fig:MSE_model_A_0}, using $M=10$ for MSE$_{av}$ and $M=10$ for $\MSEinf$ computation.
Again, note the tight match between analytical and numerical results.
For all techniques, the MSE decreases as the SNR$_m$ increases.
The MF however provides very poor performance, even for high SNR$_m$. In particular, as
SNR$_m$ tends to infinity, it shows a horizontal asymptote with $\MSEinf=\beta$.

Besides linear filtering, another technique for estimating the spectrum $\av$ is based on
interpolation~\cite{DongTong}.
The idea is to interpolate the measures $\pv$ to a regular sampling grid defined by the vector
$\xv'$ where $x'_q = (q-1)/r$, $q=1,\ldots,r$. The interpolated vector $\pv'$ is then multiplied
by the matrix $\beta\Gm_{\xv'}$. Notice that in this case $\sqrt{\beta}\Gm_{\xv'}$ is unitary i.e.
$\beta\Gm_{\xv'}\Gm_{\xv'}\Herm=\Id$, since $\xv'$ represents an equally spaced sampling.
In the figure the dashed line labeled ``Linear interp.'' shows the performance obtained using linear interpolation. The MSE has a horizontal asymptote for high SNR$_m$. While it outperforms
the MF, it clearly shows poor performance for high SNR, compared to ZF and LMMSE techniques.

Figure~\ref{fig:MSE_model_A_2} presents the performance of the LMMSE filter obtained
evaluating~(\ref{eq:MSEinf_A_LMMSE}) for different values of $\beta$,
as the SNR$_m$ varies.
In agreement with the results presented in Figure~\ref{fig:MSE_model_A_0},
the performance of the LMMSE filter
degrades as $\beta$ increases, while, as expected, it improves as
the SNR$_m$ increases.

\medskip
\example{1}{{\it We need to estimate the number of sensors required
to sample a field with $M=100$ harmonics. Each sensor provides
samples with ${\rm SNR}_m=30$\,dB.}

We choose to employ the LMMSE filter, which provides the best
performance.
Looking at Figure~\ref{fig:MSE_model_A_2}, if we allow an
$\MSEav$ of $3\cdot 10^{-3}$, then we need $\beta=0.4$, i.e., $r=(2M+1)/\beta\approx 500$ sensors.
By doubling the number of sensors ($\beta=0.2$), $\MSEav$ drops to $3\cdot 10^{-4}$.
}

\section{Analysis of Model B}
\label{sec:model_C}

Here we consider the case of sensors with jittered positions and average position,
$\xvh$, known at the sink node.
The true sensor location is: $\xv=\xvh+\deltav$, where $\deltav$ is a random vector,
as defined in Section~\ref{sec:system_model}.
The reconstruction algorithm employs the matrix $\Bm$, which is a
function of the known average positions $\xvh$. For any given $\xvh$ and
$\Bm$, similarly to~(\ref{eq:MSE}), the MSE becomes:
\begin{equation}
{\rm MSE}_\xvh = \frac{\EE_{\av,\nv,\deltav}\left[ \|\hat{\av}-\av \|^2\right]}{2M+1} = \frac{\sigma^2_a}{2M+1}\trace\{\Psim_\xvh\}
\label{eq:MSE_B}
\end{equation}
where
\begin{eqnarray}
\Psim_\xvh &=& \frac{1}{\sigma^2_a}\EE_{\av,\nv,\deltav}\left[ (\hat{\av}-\av)(\hat{\av}-\av)\Herm\right] \non
&=& \EE_{\deltav}\left[ (\Bm\Herm\Gm_\xv\Herm-\Id)(\Gm_\xv\Bm\Herm-\Id) +\alpha\Bm\Herm\Bm \right] \non
&=& \Bm\Herm\left(\EE_{\deltav}[\Gm_\xv\Herm\Gm_\xv] +\alpha\Id\right)\Bm -2\Re\left\{\EE_{\deltav}[\Gm_\xv]\Bm\right\} +\Id \non
\label{eq:Psi_B}
\end{eqnarray}
where $\Re\{\cdot\}$ represents the real part of the argument.

To proceed further we need to compute the averages over the
displacements $\deltav$, i.e.,  we need the expression of
$\EE_{\deltav}[\Gm_{\xv}]$ and
$\EE_{\deltav}\left[\Gm_\xv\Herm\Gm_\xv\right]$ as functions of
$\Gm_\xvh$, whose derivation is given in
Appendix~\ref{app:averages}. We have:
\begin{equation}\label{eq:EFx}
\EE_{\deltav}[\Gm_\xv] = \Cm\Gm_\xvh
\end{equation}
and
\begin{equation}\label{eq:EFxFhx}
\EE_{\deltav}\left[\Gm_\xv\Herm\Gm_\xv\right] =
\Gm_\xvh\Herm \Cm^2 \Gm_\xvh + \left(1-\frac{\trace\{\Cm^2\}}{2M+1}\right)\Id
\end{equation}
where $\Cm$ is a $(2M+1)\times (2M+1)$ diagonal matrix with
$(\Cm)_{kk}=C_\delta(-\jj 2\pi k)$, $k=-M,\ldots,M$, where
$C_\delta(\cdot)$ is the characteristic function of the
displacements. Under the assumption that $\deltav$ has a zero mean
Gaussian distribution we have $(\Cm)_{kk} = \exp(-2\pi^2 k^2
\sigma^2_\delta)$, $k=-M,\dots,M$.


Using~(\ref{eq:EFx}) and~(\ref{eq:EFxFhx}) in~(\ref{eq:Psi_B}) we obtain:
\begin{eqnarray}
\Psim_\xvh &=&  \Bm\Herm\left(\Gm_\xvh\Herm\Cm^2\Gm_\xvh + \gamma\Id\right)\Bm
- 2\Re\{\Cm\Gm_\xvh\Bm\} +\Id \non 
\label{eq:MSE_C2}
\end{eqnarray}
where $\gamma = 1+ \alpha +\frac{\trace\{\Cm^2\}}{2M+1}$.

In the following, in the case of the LMMSE filter\footnote{Recall that
the MF and ZF techniques, by their definition,
do not require any information on  $\sigma^2_n$ and $\sigma^2_{\delta}$}
we first consider that the variance
$\sigma^2_{\delta}$ of the sensor movement is unknown at the sink
and, hence, the sink  assumes the sensors to be fixed (i.e.,
$\deltav=\zerov$), while running the reconstruction algorithm. Then,
we consider that $\sigma^2_{\delta}$ is known and the reconstruction
algorithm employs a filter that exploits such an information to
minimize the MSE (this case is referred to as ``LMMSE for known
$\sigma_\delta^2$''.)

Finally, we remark that, while in Model A the filters used for
signal reconstruction are functions of the matrix $\Gm_\xv$ with
$\xv$ known to the sink, in Model B only the mean value of the
sensor positions $\xvh$ is known and, hence, the filters are
computed using $\Gm_\xvh$ instead of  $\Gm_\xv$.

\subsection{Matched filter}
\label{sec:MF_B}
If the sink node employs the MF in~(\ref{eq:A_B_MF})
as function of $\xvh$ (i.e., $\Bm\Herm = \beta\Gm_{\xvh}$),
then, using~(\ref{eq:MSE_C2}), we obtain:
\begin{equation}
\Psim_\xvh = \beta^2\Rm_\xvh\Cm^2\Rm_\xvh +\gamma\beta^2\Rm_\xvh -2\beta \Re\left\{\Cm\Rm_\xvh\right\}+\Id
\label{eq:MSE_C_MF}
\end{equation}
This result holds for strictly positive $\sigma^2_\delta$. Note that, for
$\sigma^2_\delta=0$ (no sensor motion), we have $\Cm=\Id$ and $\gamma=\alpha$;
thus ~(\ref{eq:MSE_C_MF}) reduces to~(\ref{eq:MSE_A_MF}).

Equation~(\ref{eq:MSE_C_MF}) refers to the MSE obtained with a given
vector $\xvh$; we are now interested in deriving the asymptotic
expression for the MSE.
Note that (\ref{eq:MSE_C_MF}) is a function of both $\Rm_\xvh$ and $\Cm$, and
contains terms of the form
$\Cm^pg(\Rm_\xvh)$ with $g(x)=1,x,x^2$ and $p=0,1,2$;
also the matrix $\Rm_\xvh$ depends on $M$ and $r$, while the matrix
$\Cm$ depends on $M$ and $\sigma^2_\delta$.
The definition of the asymptotic MSE in~(\ref{eq:MSEinf}) refers to
the case where the number of harmonics $M$ and the number of sensors $r$ grow
to infinity with constant ratio $\beta$; if this is directly applied
to~(\ref{eq:MSE_C_MF}), information losses may arise. Indeed, we
have:
\begin{eqnarray}
\hspace{-6mm} \phi(\Cm^p) \hspace{-3mm} &= &\hspace{-3mm} \limbeta \frac{1}{2M+1}\trace\{\Cm^p\}\non
\hspace{-3mm} & =&\hspace{-3mm} \lim_{M\rightarrow +\infty}  \frac{1}{2M+1}
\sum_{k=-M}^M \ee^{-2p\pi^2 k^2 \sigma^2_\delta}=0
\end{eqnarray}
and thus all terms depending on the matrix $\Cm$ would vanish
regardless of the value of $\sigma^2_\delta$. On the contrary, in a
realistic situation we expect to obtain high reconstruction quality
when the standard deviation of the motion ($\sigma_\delta$) is
smaller than or comparable to the average sensor separation ($1/r$),
and a significant degradation of the reconstruction quality when
$\sigma_\delta$ is much larger than the average sensor separation.
To distinguish such different conditions, we define the {\em
signal-to-noise ratio on the motion} as:
\[ {\rm SNR}_{x} =\frac{(1/r)^2}{\sigma^2_\delta} = \frac{1}{\omega^2}\]
where $\omega=\sigma_\delta r$.
We then redefine the asymptotic MSE as the limit of the average MSE
for $M,r\rightarrow+\infty$, with constant $\beta=(2M+1)/r$ {\em and} constant
$\omega=\sigma_{\delta} r$.
In this case,
\begin{eqnarray}
\phi(\Cm^p)
\hspace{-3mm}&=& \hspace{-3mm}\limbetaomega \frac{1}{2M+1} \sum_{k=-M}^M \exp\left(-2p\pi^2 k^2 \sigma^2_\delta\right) \non
\hspace{-3mm}&=& \hspace{-3mm}\int_{-1/2}^{1/2} \exp\left(-2p\pi^2 z^2 \beta^2\omega^2\right) \dd z \non
\hspace{-3mm}&=& \hspace{-3mm}\sqrt{\frac{\pi}{4}}\frac{\erf\left(\sqrt{\frac{p}{2}}\pi\beta\omega\right)}{\sqrt{\frac{p}{2}}\pi\beta\omega}
=\nu\left(\sqrt{\frac{p}{2}} \beta\omega\right)
\label{eq:limit_Mp}
\end{eqnarray}
where $\nu(x)=\sqrt{\pi/4}\,\erf(\pi x)/(\pi x)$. Notice that $\nu(0)=1$ and
$\lim_{x\rightarrow +\infty}\nu(x) = 0$.
Also,  we have:
\begin{eqnarray} \label{eq:Psi-gamma}
\phi(\gamma) \hspace{-3mm} & =  &\hspace{-3mm} 1+ \alpha - \limbetaomega \frac{\trace\{\Cm^2\}}{2M+1}
= 1+\alpha -\nu(\beta\omega)
\end{eqnarray}

Using the new definition and~(\ref{eq:MSE_C_MF}), the asymptotic expression of the MSE
becomes:
\begin{eqnarray}
\MSEinf
&=& \trace\left\{\Psim_\xvh\right\}\non
&=& \phi\left(\beta^2\Rm_\xvh\Cm^2\Rm_\xvh +\gamma\beta^2\Rm_\xvh -2\beta \Re\left\{\Cm\Rm_\xvh\right\}+\Id\right)\non
&=& \beta^2 \phi\left(\Cm^2\Rm_\xvh^2\right)+\beta^2\phi\left(\gamma\Rm_\xvh\right)-2\beta\phi\left(\Cm\Rm_\xvh\right) +1\non
&=& \beta^2\phi(\Cm^2)\phi(\Rm_\xvh^2)+\beta^2\phi(\gamma)\phi(\Rm_\xvh) \non
& & \qquad -2\beta\phi(\Cm)\phi(\Rm_\xvh)  + 1\non
&=& \nu(\beta\omega)\EE[\lambda^2] +\beta\phi(\gamma)\EE[\lambda] \non
& & \qquad -2\nu(\beta\omega/\sqrt{2})\EE[\lambda] +1 \non
&=& \nu(\beta\omega)(1+\beta) +\beta(1+\alpha-\nu(\beta\omega)) \non
& & \qquad -2\nu(\beta\omega/\sqrt{2}) +1 \non
&=& \beta(1+\alpha) +\nu(\beta\omega) -2\nu(\beta\omega/\sqrt{2}) +1
\label{eq:MSEinf_C_MF}
\end{eqnarray}
Here we used the following facts:
\begin{itemize}
\item $\phi(\Re\{\Cm\Rm_\xv\}) = \phi(\Cm\Rm_\xv)$ since $\Rm_\xvh$ is Hermitian and $\Cm$ is real and diagonal;
\item $\trace\{\Xm_1\Xm_2\} = \trace\{\Xm_2\Xm_1\}$ for any square matrix $\Xm_1$ and $\Xm_2$;
\item $\phi(\Cm^p\Rm_\xvh^q) = \phi(\Cm^p)\phi(\Rm_\xvh^q)$ for any positive integer $p$ and $q$.
This assumption holds only if $\Cm$ and $\Rm_\xvh$ are asymptotically free~\cite{TulinoVerdu}.
Since asymptotical freeness is in general very hard to prove, we will simply verify the validity
of such assumption through numerical results.
\item $\EE[\lambda^2] =1+\beta$ and $\EE[\lambda] =1$ (see~\cite{NordioChiasseriniViterbo});
\end{itemize}
Equation~(\ref{eq:MSEinf_C_MF}) reduces to~(\ref{eq:MSEinf_A_MF2}) for $\omega=0$,
while it reduces to $\MSEinf=1+\beta(1+\alpha)$ for $\omega=+\infty$.


\subsection{ZF filter}

In this case the sink node employs the ZF filter in (\ref{eq:A_B_ZF}) but,
knowing only the average value of the sensor positions, the filter
results to be a function of $\xvh$: $\Bm\Herm = \Rm_\xvh^{-1}\Gm_\xvh$,
and the matrix $\Psim_\xvh$ can be written as:
\begin{equation}
\Psim_\xvh = \gamma \Rm_\xvh^{-1} + (\Cm-\Id)^2
\label{eq:MSE_C_ZF}
\end{equation}
We observe that, when $\sigma^2_\delta=0$ (no sensor motion), we have $\Cm=\Id$ and
$\gamma=\alpha$, thus~(\ref{eq:MSE_C_ZF}) reduces to~(\ref{eq:MSE_A_ZF}).

Using (\ref{eq:limit_Mp}) and (\ref{eq:Psi-gamma}),  the asymptotic MSE is:
\begin{eqnarray}
\MSEinf &=& \phi\left(\gamma \Rm_\xvh^{-1} + (\Cm-\Id)^2 \right) \non
&=&  \beta\left(1+\alpha-\nu(\beta\omega)\right)\EE\left[\frac{1}{\lambda}\right] \non
& & \qquad +1+\nu(\beta\omega) -2\nu(\beta\omega/\sqrt{2})
\label{eq:MSEinf_C_ZF}
\end{eqnarray}
Equation~(\ref{eq:MSEinf_C_ZF}) reduces to~(\ref{eq:MSEinf_A_ZF}) for $\omega=0$,
while it reduces to $\MSEinf=1+\beta(1+\alpha)\EE[1/\lambda]$ for $\omega=+\infty$.

\subsection{LMMSE filter neglecting $\sigma_{\delta}^2$  \label{subsec:neglecting-sigma_d}}

If the sink employs the filter in~(\ref{eq:A_B_LMMSE}) computed
using $\xvh$ (i.e., $\Bm\Herm = \Am^{-1}_\xvh\Gm_\xvh$, where $\Am_\xvh= \Rm_\xvh+\alpha\Id$), then the matrix $\Psim_\xvh$
in (\ref{eq:MSE_C2}) becomes:
\begin{equation}
\Psim_\xvh = \Am^{-1}_\xvh\Rm_\xvh(\Cm^2\Rm_\xvh +\gamma\Id)\Am^{-1}_\xvh -2\Re\{\Cm\Rm_\xvh\Am^{-1}_\xvh\} +\Id
\label{eq:MSE_C_LMMSE}
\end{equation}
For $\sigma^2_\delta=0$ (i.e., $\Cm=\Id$ and $\gamma=\alpha$),
(\ref{eq:MSE_C_LMMSE}) reduces to~(\ref{eq:MSE_A_LMMSE}).

Using the properties described in Section~\ref{sec:MF_B} the asymptotic MSE is:
\begin{eqnarray}
\MSEinf  &=& 1+\left(\nu(\beta\omega)-2\nu(\beta\omega/\sqrt{2})\right)
                 \EE\left[\frac{\lambda^2}{(\lambda+\alpha\beta)^2}\right]\non
 &\hspace{-18mm}&\hspace{-18mm}+\beta\left(1+\alpha-\nu(\beta\omega)
-2\alpha\nu(\beta\omega/\sqrt{2})\right)\EE\left[\frac{\lambda}{(\lambda+\alpha\beta)^2}\right]
\label{eq:MSEinf_C_LMMSE}
\end{eqnarray}
Equation~(\ref{eq:MSEinf_C_LMMSE}) reduces to~(\ref{eq:MSEinf_A_LMMSE}) for $\omega=0$,
while it becomes: $\MSEinf=1+\beta(1+\alpha)\EE[\lambda/(\lambda+\alpha\beta)^2]$  for $\omega=+\infty$.

\subsection{LMMSE filter for known $\sigma^2_\delta$ \label{subsec:known-sigma_d}}
We now consider the linear LMMSE filter optimized for the case where $\sigma^2_\delta$ is
known at the sink. We find the optimal $\Bm$
minimizing $\trace\{\Psim_\xvh\}$; that is, we null the derivative
of~(\ref{eq:MSE_B}) with respect to $\Bm$. We employ the following properties that hold
for any square matrix $\Xm$ \cite{MatrixBook}:
\begin{eqnarray}
  \frac{\partial}{\partial \Bm} \Re\trace\left\{\Xm\Bm\right\} \hspace{-3mm}&=& \hspace{-3mm}\Xm\Herm\non
  \frac{\partial}{\partial \Bm} \trace\left\{\Bm\Herm\Xm\Bm\right\} \hspace{-3mm}&=& \hspace{-3mm}2\Xm\Bm \quad\quad
\mbox{if}\;\; \Xm = \Xm\Herm \nonumber
\end{eqnarray}
Then, we have:
\[ \frac{\partial {\rm MSE}_\xvh(\Bm)}{\partial \Bm} =
\frac{2\sigma^2_a\left(\Gm_\xvh\Herm\Cm^2\Gm_\xvh+\gamma\Id\right)\Bm}{2M+1}
-\frac{2\sigma^2_a\Gm_\xvh\Herm\Cm}{2M+1}=\zerov \]
Solving for $\Bm$, we obtain the expression of the LMMSE filter
\begin{equation}
\Bm\Herm =  \left(\Cm\Rm_\xvh\Cm+\gamma\Id\right)^{-1}\Cm\Gm_\xvh
\label{eq:C_B_LMMSE_2}
\end{equation}
Substituting~(\ref{eq:C_B_LMMSE_2}) into~(\ref{eq:MSE_C2}), we have:
\begin{equation}
\Psim_\xvh = \gamma \left(\Cm\Rm_\xvh\Cm +\gamma\,\Id\right)^{-1}
\label{eq:MSE_C_LMMSE2}
\end{equation}
In this case an explicit expression of $\MSEinf$ is hard to obtain.
However, we were able to find the following {\em lower bound} that
turns out to be very tight, as shown by the results presented in
the following section
\begin{eqnarray}
\label{eq:MSEinf_C_LMMSE-D}
\MSEinf &=& \phi(\gamma (\Cm\Rm_\xvh\Cm +\gamma\,\Id)^{-1})\nonumber\\
& \geq &\frac{1}{\phi\left(\frac{1}{\gamma}(\Cm\Rm_\xvh\Cm +\gamma\,\Id)\right)}\non
& = &\frac{\phi(\gamma)}{\phi\left(\Cm^2\Rm_\xvh\right) +\phi(\gamma)}\non
& = & \beta\frac{1+\alpha-\nu(\beta \omega)}{\beta(1+\alpha)+\nu(\beta \omega)(1-\beta)}
\end{eqnarray}
where to derive the last expression we exploited (\ref{eq:limit_Mp}), (\ref{eq:Psi-gamma}),
(\ref{eq:g(X)}) and the fact that $\EE[\lambda]=1$.

\insertfig{1.00}{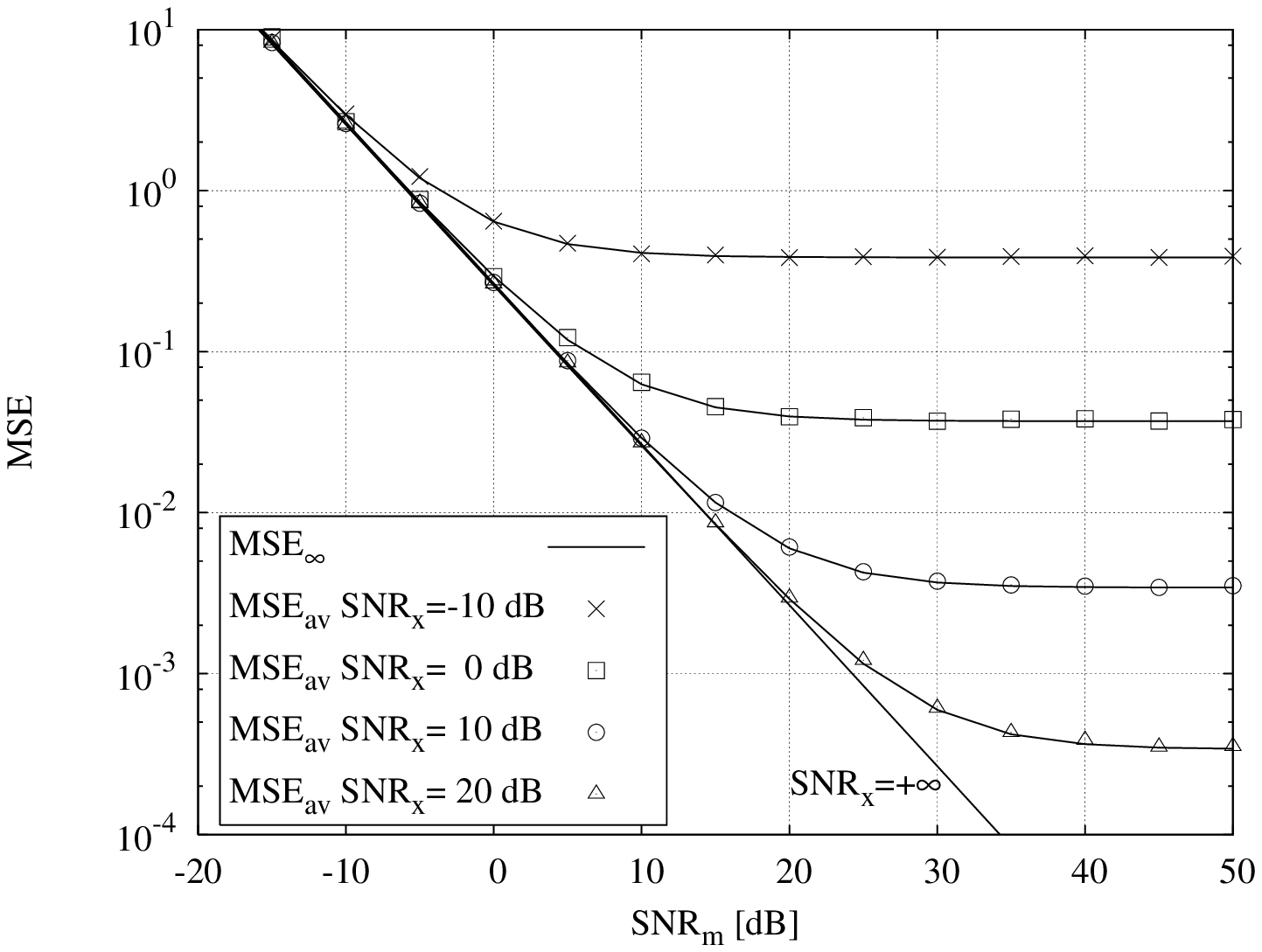}{Performance of the ZF filter for $\beta=0.2$ and $M=10$, 
when $\sigma^2_\delta$ is neglected}{fig:MSE_model_B_0}
\insertfig{1.00}{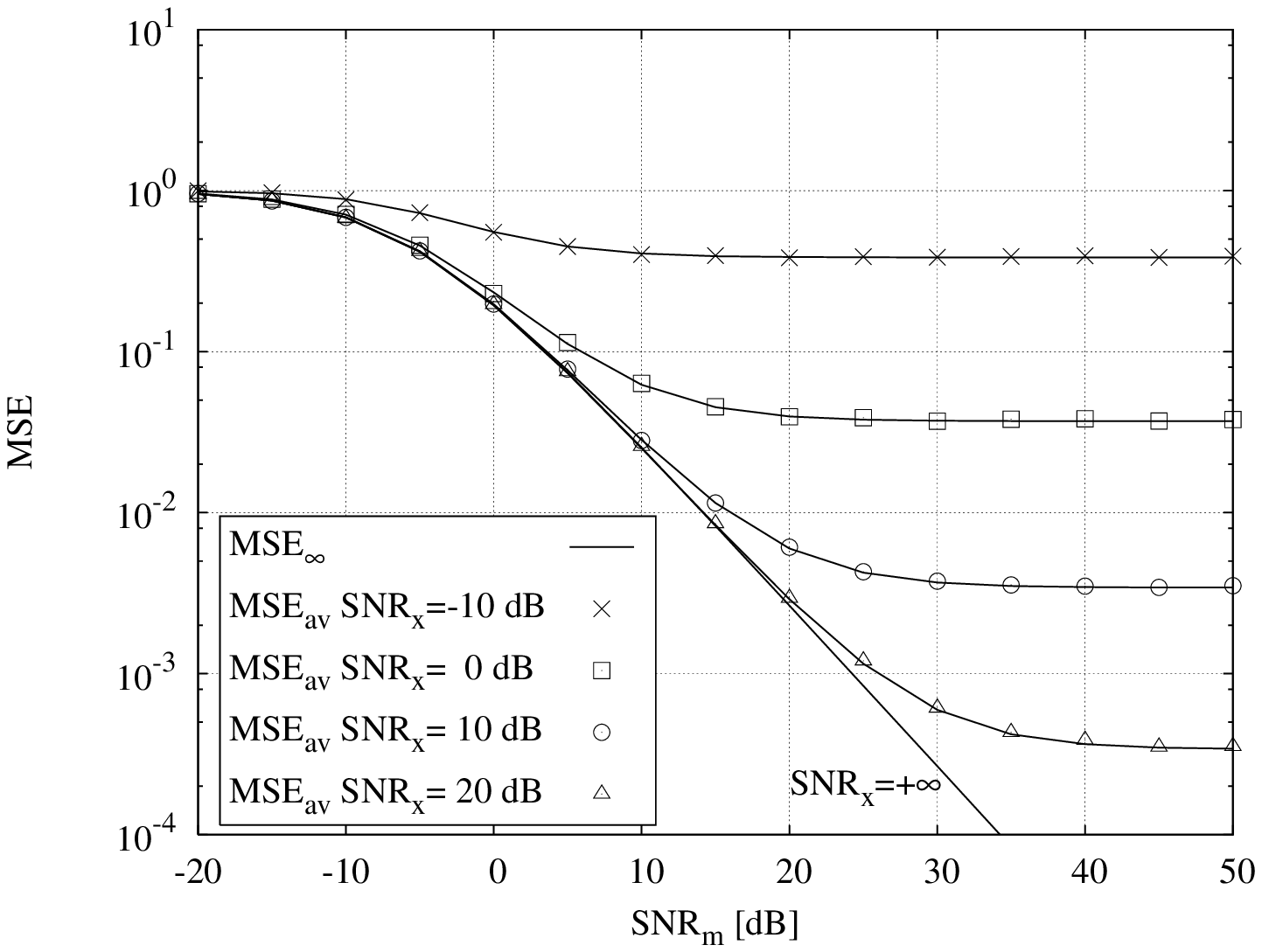}{Performance of the LMMSE filter~(\ref{eq:A_B_LMMSE}) versus SNR$_m$, 
for $\beta=0.2$ and $M=10$, when $\sigma^2_\delta$ is neglected}{fig:MSE_model_B_1}
\insertfig{1.00}{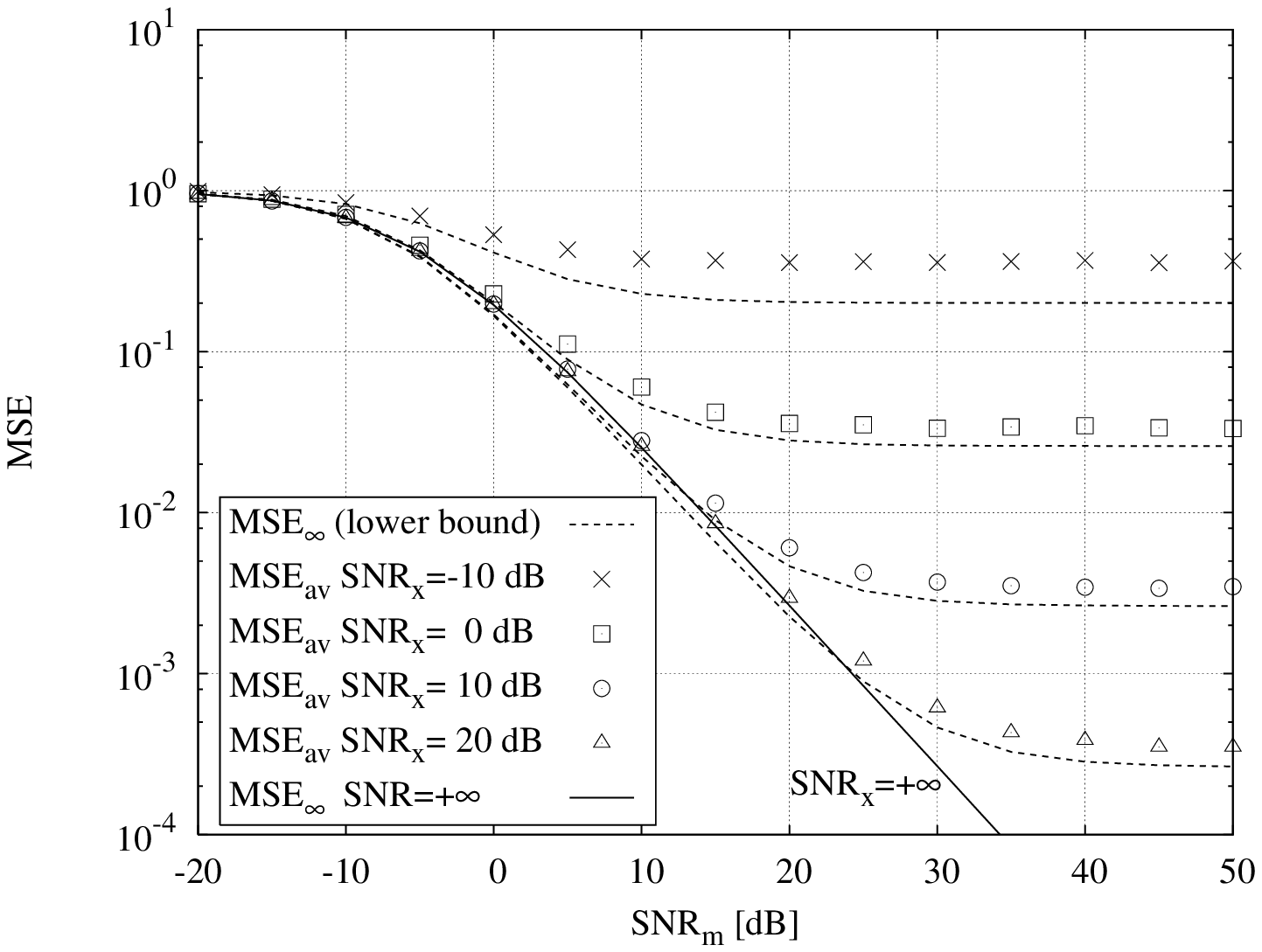}{Performance of the LMMSE filter~(\ref{eq:C_B_LMMSE_2}) with perfect
knowledge of $\sigma^2_\delta$ versus SNR$_m$, for $\beta=0.2$ and $M=10$}{fig:MSE_model_B_2}
\insertfig{1.00}{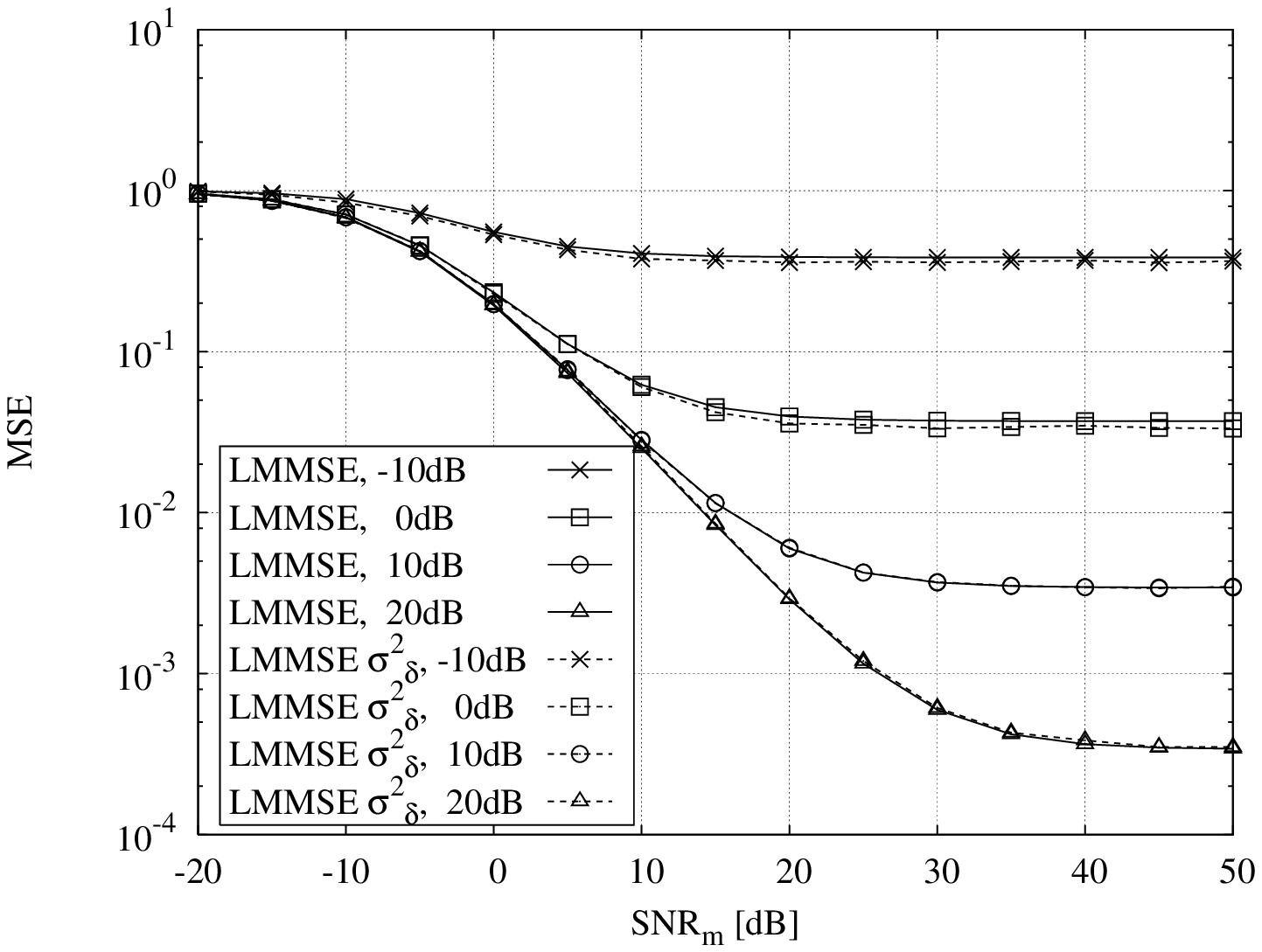}{Performance comparison of the LMMSE filter neglecting
$\sigma^2_\delta$~(\ref{eq:A_B_LMMSE})
against the LMMSE filter with perfect knowledge of $\sigma^2_\delta$, as SNR$_m$ varies and
for $\beta=0.2$ and $M=10$}{fig:MSE_model_B_3}
\insertfig{1.00}{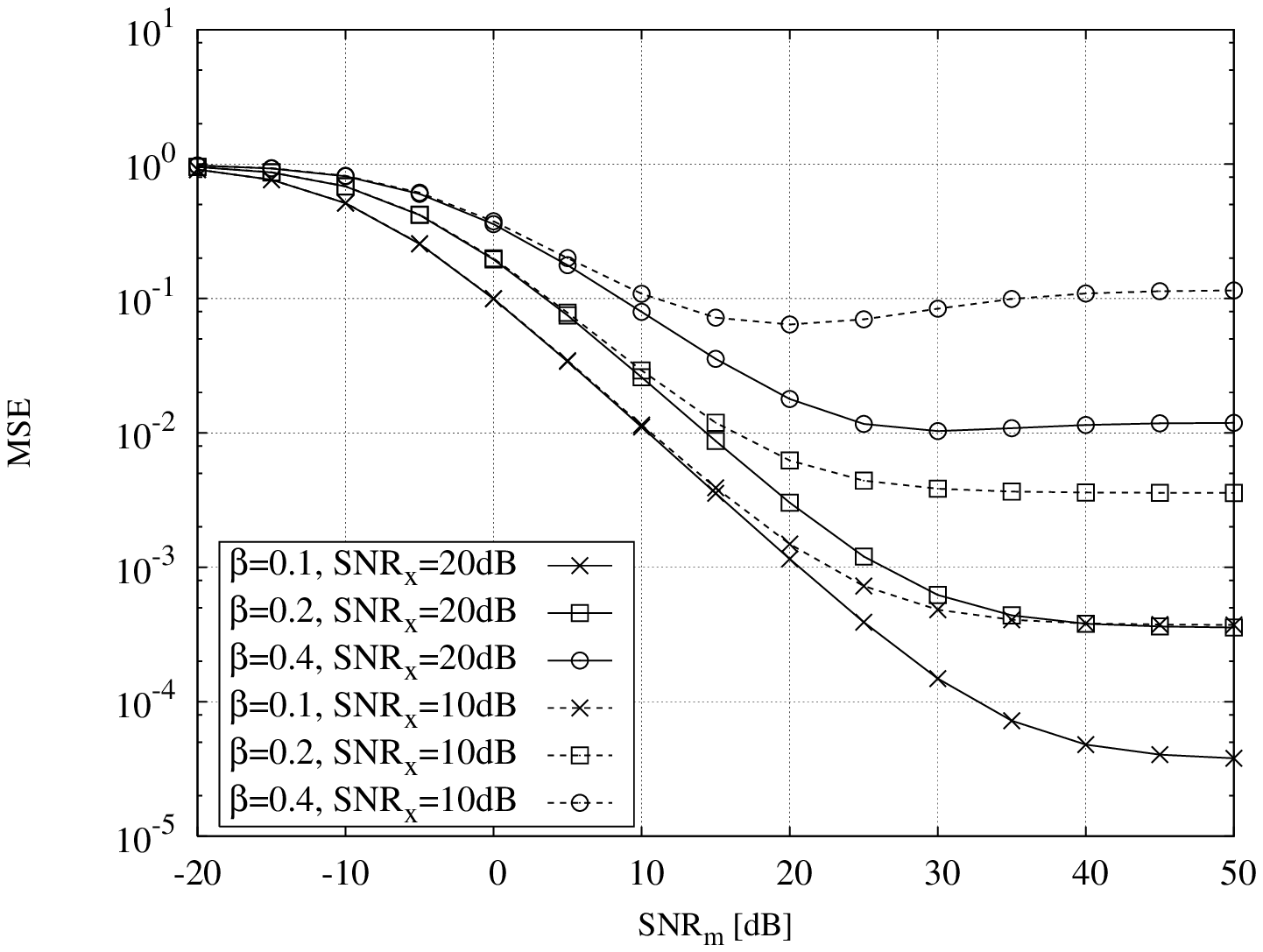}{Performance of the LMMSE filter when $\sigma^2_\delta$
is neglected as SNR$_m$ varies, for SNR$_x=10,20$~dB, and $\beta=0.1,0.2,0.4$ and $M=10$}{fig:MSE_model_B_4}

\subsection{Results \label{subsec:C_numerical_results}}

We now show the performance of the filters analyzed under Model B.
Regarding the ZF filter (\ref{eq:A_B_ZF}),
Figure~\ref{fig:MSE_model_B_0} compares the asymptotic MSE evaluated
through~(\ref{eq:MSEinf_C_ZF}) (represented by solid lines and
labeled by ``$\MSEinf$'') against the average MSE (represented by
points and labeled by ``$\MSEav$''). The $\MSEav$ is obtained by
generating $100$ realizations of the measures as
in~(\ref{eq:p_fixed}), with $M=10$, computing the estimates as
in~(\ref{eq:hat_a}) and averaging the square error
$\|\av-\hat{\av}\|^2$. The MSE is shown in the log scale plotted
versus ${\rm SNR}_m$, for $\beta=0.2$ and different values of ${\rm
SNR}_x$.

Similarly, Figure~\ref{fig:MSE_model_B_1} presents the performance of the
LMMSE filter~(\ref{eq:A_B_LMMSE}). Here the curves labeled by ``$\MSEinf$'', generated through evaluation of~(\ref{eq:MSEinf_C_LMMSE}),
and the points in the plot, labeled by ``$\MSEav$'', have been obtained as for Figure~\ref{fig:MSE_model_B_0}.

In both the plots the solid line labeled by ``SNR$_x=+\infty$'' refers to the case
where $\omega=0$, i.e. $\deltav=0$, and correspond to the performance provided by Model A under the same conditions.

 The excellent match between the asymptotic results and the
numerical simulation confirms the validity of the asymptotic analysis as an effective tool
to characterize the performance of the reconstruction techniques.

Also, comparing Figures~\ref{fig:MSE_model_B_0} and \ref{fig:MSE_model_B_1}, we observe that
the performances of the ZF and the LMMSE filters are similar for SNR$_m>10$~dB for
any value of ${\rm SNR}_x$, while, for lower SNR$_m$,
the LMMSE filter outperforms the ZF filter.

Figure~\ref{fig:MSE_model_B_2} compares the performance of the LMMSE filter~(\ref{eq:C_B_LMMSE_2}), which
has knowledge of $\sigma^2_\delta$, with its lower bound~(\ref{eq:MSEinf_C_LMMSE-D}) (dashed lines), as SNR$_m$ varies.
We consider $\beta=0.2$ and different values of ${\rm SNR}_x$. Notice that the lower bound is
very tight, especially for high values of ${\rm SNR}_x$.
The points in the plot, labeled by ``$\MSEav$'' have been obtained as for Figure~\ref{fig:MSE_model_B_0}, using $M=10$.
Here, as well as in Figure~\ref{fig:MSE_model_B_1}, the the case ${\rm SNR}_x=\infty$ (solid line) is shown, and
corresponds to the performance of the LMMSE filter for signal model A. Indeed, for ${\rm SNR}_x=\infty$ (i.e., $\sigma^2_\delta=0$ and $\deltav=0$), we have $\Cm=\Id$ and $\gamma=\alpha$, and~(\ref{eq:MSE_C_LMMSE2}) simplifies to~(\ref{eq:MSE_A_LMMSE}).

Figure~\ref{fig:MSE_model_B_3} compares the performance of the LMMSE filter (\ref{eq:A_B_LMMSE}),
labeled by ``LMMSE'' (solid lines), and of
the LMMSE filter~(\ref{eq:C_B_LMMSE_2}), labeled by ``LMMSE $\sigma^2_\delta$'' (dashed lines),
for the same parameter setting as in Figure~\ref{fig:MSE_model_B_2}.
For the considered value of $\beta$ ($\beta$=0.2), the filter in (\ref{eq:C_B_LMMSE_2})
outperforms the simpler filter~(\ref{eq:A_B_LMMSE}) for any value of SNR$_m$ and SNR$_x$,
but the performance gain is always negligible.

Figure~\ref{fig:MSE_model_B_4} shows the performance of the
LMMSE filter~(\ref{eq:A_B_LMMSE}) neglecting $\sigma^2_\delta$,
obtained through evaluation of~(\ref{eq:MSEinf_C_LMMSE}) for
SNR$_x=10$\,dB (dashed lines) and SNR$_x=20$\,dB (solid lines), and
for $\beta=0.1,0.2,0.4$.
While the $\MSEinf$ of the LMMSE filter~(\ref{eq:A_B_LMMSE}) always tends to 1 for
small values of ${\rm SNR}_m$ (i.e., large values of $\alpha$),
for high ${\rm SNR}_m$ (i.e., low $\alpha$) its behavior depends on $\beta$.
Indeed the term $\EE[\lambda/(\lambda+\alpha\beta)^2)]$ on the right hand side
of~(\ref{eq:MSEinf_C_LMMSE}) reduces to $\EE[1/\lambda]$ for $\alpha\rightarrow 0$.
As explained in Section~\ref{sec:A_ZF}, $\EE[1/\lambda]$ diverges for
$\beta>\beta^\star\approx0.35$ and so the MSE (see the lines with $\circ$ markers in the plot).
This behavior is more evident as $\beta$ increases and the MSE is large,
for any SNR$_m$. These results, however, are of no interest from the application point
of view since a system characterized by such poor performance is not working.

\insertfig{1.00}{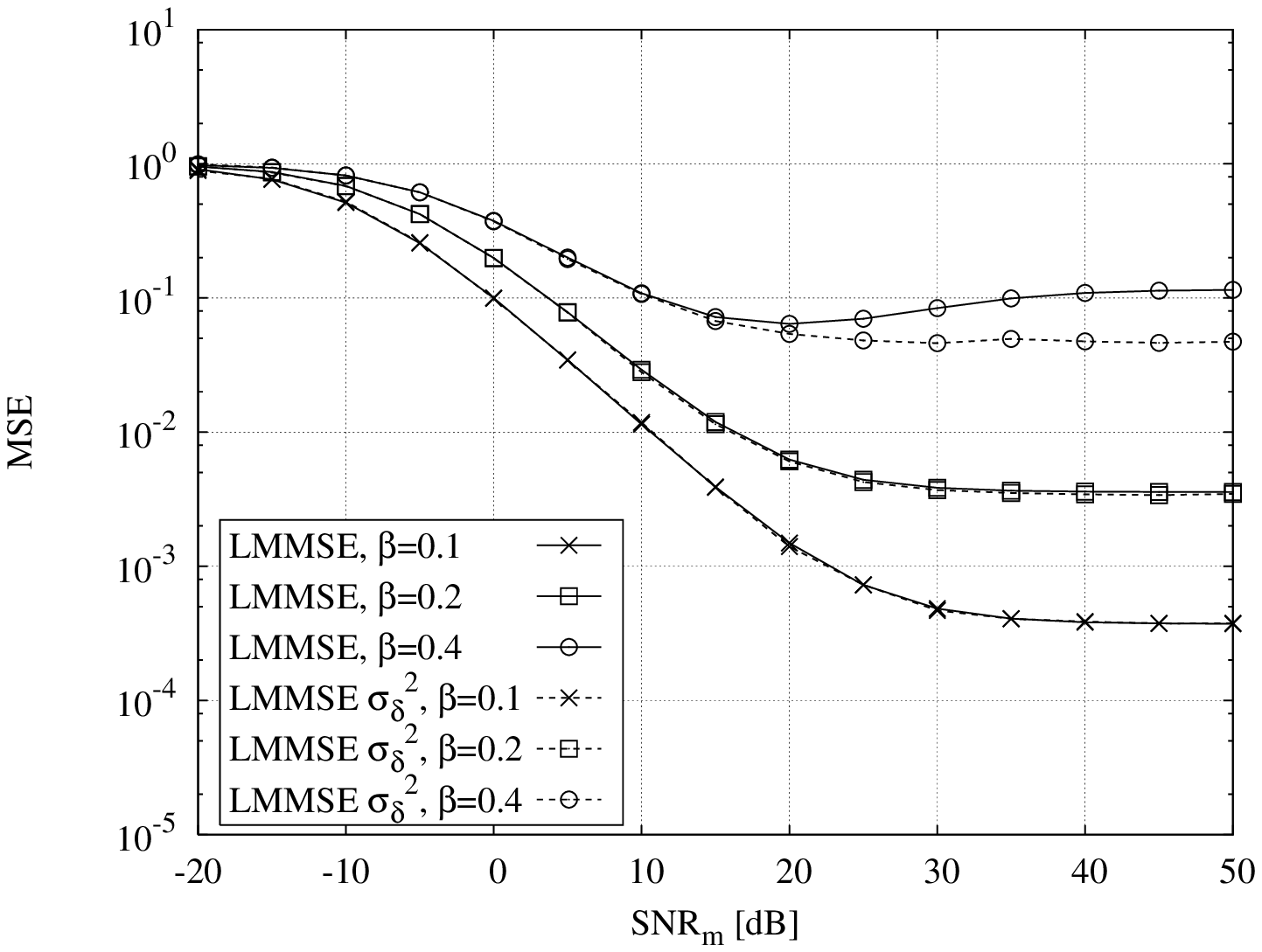}{Performance comparison of the LMMSE filter neglecting $\sigma^2_\delta$ against
the LMMSE filter with perfect knowledge of $\sigma^2_\delta$, as SNR$_m$ varies, for SNR$_x=10$~dB, and $M=10$}
{fig:MSE_model_B_5}

Finally, Figure~\ref{fig:MSE_model_B_5} compares
the performance of the LMMSE filter~(\ref{eq:C_B_LMMSE_2})
labeled by ``LMMSE $\sigma^2_\delta$'' (dashed lines) and the performance of the LMMSE filter~(\ref{eq:A_B_LMMSE}),
labeled by ``LMMSE'' (solid lines), for SNR$_x=10$\,dB and $\beta=0.1,0.2,0.4$.

In general the filter~(\ref{eq:C_B_LMMSE_2}) performs always better than filter~(\ref{eq:A_B_LMMSE}).
In particular, for $\beta<\beta^\star$ the two filters show very similar performance, while,
when $\beta>\beta^\star$,
the filter~(\ref{eq:C_B_LMMSE_2}) does not diverge for high SNR$_m$.
This is shown in Figure~\ref{fig:MSE_model_B_5},
where, for $\beta>0.35$ and high values of ${\rm SNR}_m$,
the advantage of exploiting the knowledge of $\sigma^2_\delta$
becomes evident.

\medskip
\example{2}{{\it Consider $r=1000$ buoys  deployed in water and
equipped with sensors, which provide noisy measures with ${\rm SNR}_m=30$\,dB.
Buoys are moving but the variance $\sigma^2_\delta=10^{-7}$ is unknown to the reconstruction algorithm.
We need to estimate the maximum number of harmonics of the field that the network can sample and reconstruct
with an average MSE lower than $5~10^{-3}$.
}

Since  ${\rm SNR}_m$ is known to the reconstruction algorithm
while $\sigma^2_\delta$ is not, we employ the LMMSE filter in Sec.~\ref{subsec:neglecting-sigma_d}.
We have: ${\rm SNR}_x=1/(\sigma^2_\delta r^2)=10$. Looking at
Figure~\ref{fig:MSE_model_B_5}, we notice that, for ${\rm
SNR}_m=30$\,dB, values of $\MSEav$ lower than $5~10^{-3}$ can be
obtained only for $\beta<0.2$. The maximum number of harmonics is
then $M=(r\beta-1)/2\approx 100$.
} 

\medskip
\example{3}{ {\it Consider a  network of sensor with jittered positions
characterized by $\beta$=0.4 and {\rm SNR}$_x=10$\,dB, and assume that
these values are known to the reconstruction algorithm. We want to
determine which type of sensor devices should be used  in order to
minimize the $\MSEav$. In other words, we ask ourselves how accurate
the sensor measurements need to be (clearly, more expensive devices
provide a higher {\rm SNR}$_m$). }

Since SNR$_x$=10\,dB is known to the reconstruction algorithm, we
can employ the LMMSE filter given in (\ref{eq:C_B_LMMSE_2}). Looking
at Figure~\ref{fig:MSE_model_B_5}, we notice that the performance of
the filter for $\beta=0.4$ shows a horizontal asymptote
corresponding to an average MSE of $5\cdot 10^{-2}$. Thus, an SNR$_m
= 25$\,dB is enough to achieve the best performance.
} 

\section{Summary of Results}
\label{sec:summary}

Our main results for the system models A and B are as follows.
\begin{description}
\item[{\bf Model A}] \hspace{0.7cm} (fixed sensors and noisy measures):
\begin{itemize}
\item  for a given $\beta$, the MSE provided by any of the reconstruction techniques
is lower bounded by $\alpha\beta/(1+\alpha\beta)$ and worsen with increasing
$\beta$ (i.e., the ratio of the number of harmonics
 to the number of sampling sensors);
the MF in~(\ref{eq:A_B_MF}) is the only filter which
does not require matrix inversion,
however it provides poor performance in all of the considered cases;
\item the ZF filter
provides high quality performance only
for high ${\rm SNR}_m$ (namely, ${\rm SNR}_m>10$\,dB) and $\beta<0.35$;
\item the performance of the LMMSE filter, instead,
is good moderate values of ${\rm SNR}_m$ and $\beta<1$.
\end{itemize}

\item[{\bf  Model B}] \hspace{0.7cm} (sensors with jittered positions and noisy measures):
\begin{itemize}
\item for a given $\beta$ the MSE provided by any of the reconstruction techniques
is lower bounded by (\ref{eq:MSEinf_C_LMMSE-D});
\item the performance of all reconstruction techniques worsen with
increasing $\beta$ and ${\rm SNR}_x$;
\item the advantage of exploiting the knowledge of ${\rm SNR}_x$
in the filter design is negligible for low $\beta$ and low ${\rm SNR}_m$,
while it is of fundamental importance to obtain a high
quality reconstruction for $\beta>0.35$ and large values of ${\rm SNR}_m$.

\end{itemize}
\end{description}

\section{Conclusions\label{sec:conclusions}}
We addressed the problem of reconstructing band-limited fields from
measurements taken by irregularly deployed sensors, and we studied
the effects of noisy measures and jittered sensors positions on the reconstruction quality.
We analytically derived
the performance of several linear filters in terms of
the MSE of the field estimates. We also studied
the asymptotic MSE, obtained as the number of harmonics and the number of
sensors grow to infinity while their ratio $\beta$ is kept constant.
We found that the asymptotic analysis is an effective tool to characterize the
performance of the reconstruction techniques
even for a small number of sensors, and we investigated the impact
that the parameter $\beta$ has on the system performance.
In~\cite{NordioChiasseriniViterbo} we observed that random sampling without any type of noise
would require more than twice the sampling rate ($\beta<0.5$) of minimum regular sampling ($\beta=1$)
to get a reliable reconstruction (without ill conditioning problems) with high probability.
The number of sensors further increases ($\beta<0.2$) compared to regular sampling when measurement
noise (model A) and sensors position jitter (model B) are present.


\appendices

\section{Proof of~(\ref{eq:g(X)})}
\label{app:proof3}
Let $\Xm = \Am\Am\Herm$ be an Hermitian matrix where $\Am$ is an arbitrary random matrix.
Let $g(\cdot)$ be an analytic function, defined for $x\in \RR^+$ that can be written as:
$g(x) = \sum_{i=0}^{+\infty} a_i x^i $, with finite coefficients $a_i$.
Considering that $\phi(\Xm^p)$ is the $p$-th moment of the asymptotic eigenvalue
distribution of $\Xm$, i.e., $\phi(\Xm^p) = \EE[\xi^p]$
where $\xi$ is the random variable distributed as the asymptotic eigenvalues of $\Xm$,
and the continuity of the function $\phi(\cdot)$, we have:
\begin{eqnarray*}
\phi(g(\Xm))
&=&
\phi\left(\sum_{i=0}^{+\infty} a_i\Xm^i\right)
\non
&=&
\sum_{i=0}^{+\infty} a_i \phi(\Xm^i)
=\sum_{i=0}^{+\infty} a_i \EE[\xi^i]
\non
&=&
\EE\left[\sum_{i=0}^{+\infty} a_i \xi^i \right]
=\EE\left[g(\xi)\right]
\end{eqnarray*}

\section{Proof of Lemma~\ref{lemma:1}}
\label{app:proof}
Using~(\ref{eq:theorem_equation}) the $kq$-th entry of $\Gm_\xv$ is
\begin{eqnarray*}
(\Gm_\xv)_{kq}
&=& \sum_{n=0}^\infty \frac{1}{n!} (\Wm^n \Gm_\xvh \Deltam^n)_{kq} \non
&=& \sum_{n=0}^\infty \frac{1}{n!} \sum_h\sum_j (\Wm^n)_{kh} (\Gm_\xvh)_{hj} (\Deltam)_{jq}^n \non
&=& \sum_{n=0}^\infty \frac{1}{n!} (\Wm^n)_{kk} (\Gm_\xvh)_{kq} (\Deltam^n)_{qq} \non
&=& \exp\left((\Wm)_{kk}(\Deltam)_{qq} \right) (\Gm_\xvh)_{kq} \non
&=& \exp\left(-\jj 2\pi k \delta_q\right) (\Gm_\xvh)_{kq} \non
&=& \frac{1}{\sqrt{2M+1}}\exp\left(-\jj 2\pi k \delta_q\right) \exp\left(-\jj 2\pi k \hat{x}_q\right)\non
&=& \frac{1}{\sqrt{2M+1}}\exp\left(-\jj 2\pi k x_q\right) \nonumber
\end{eqnarray*}
which matches its definition.

\section{Computation of $\EE \left[\Gm_\xv\right]$ and
$\EE\left[\Gm_\xv\Herm\Gm_\xv\right]$ as functions of $\Gm_\xvh$}
\label{app:averages}

We derive here the expressions of $\EE \left[\Gm_\xv\right]$ and
$\EE\left[\Gm_\xv\Herm\Gm_\xv\right]$ as functions of $\Gm_\xvh$.
\paragraph{Computation of $\EE \left[\Gm_\xv\right]$}
Using Lemma~\ref{lemma:1} we have
\[ \EE \left[\Gm_\xv\right] = \EE\left[\sum_{n=0}^\infty \frac{1}{n!}\Wm^n \Gm_\xvh \Deltam^n\right]
= \sum_{n=0}^\infty \frac{1}{n!} \Wm^n \Gm_\xvh\EE\left[\Deltam^n\right] \]
The average of $\Deltam^n$ is given by:
$\EE\left[\Deltam^n\right]=\int_{-\infty}^{+\infty} x^n f_\delta(x) \dd x\,\Id_r = \mu^{(n)}\,\Id_r$,
where $\Id_r$ is the $r\times r$ identity matrix and $ \mu^{(n)}$
is the $n$-th moment of $\delta$. Hence,
\begin{eqnarray}
\EE \left[\Gm_\xv\right] &=& \sum_{n=0}^\infty \frac{1}{n!}\Wm^n\Gm_\xvh \int_{-\infty}^{+\infty}x^n f_\delta(x) \dd x \non
&=& \int_{-\infty}^{+\infty}\sum_{n=0}^\infty \frac{x^n}{n!}\Wm^n\,f_\delta(x)\dd x\,\Gm_\xvh \non
&=& \int_{-\infty}^{+\infty} \exp(x\Wm) f_\delta(x) \dd x\,\Gm_\xvh = \Cm\Gm_\xvh \nonumber
\end{eqnarray}
where\footnote{Let $\Am=\diag(a_1,\ldots,a_n)$ be a diagonal $n\times n$ matrix.
The exponential of $\Am$, denoted by $\exp(\Am)$, is the diagonal matrix whose elements
are $[\exp(a_1),\ldots, \exp(a_n)]$.}
\begin{eqnarray*}
\Cm &=& \int_{-\infty}^{+\infty}\sum_{n=0}^\infty \frac{x^n}{n!}\Wm^n\,f_\delta(x)\dd x\non
&=&\sum_{n=0}^\infty\frac{\mu^{(n)}}{n!}\Wm^n = \int_{-\infty}^{+\infty} \exp(x\Wm) f_\delta(x) \dd x
\end{eqnarray*}
is a $(2M+1)\times (2M+1)$ diagonal matrix and
\[ (\Cm)_{kk} = \int_{-\infty}^{+\infty} \exp(-\jj 2\pi k x) f_\delta(x) \dd x \]
is the {\em characteristic function} of the random variable $\delta$, $C_\delta(s)$,
sampled in $s=-\jj 2\pi k$.
In particular when $\delta$ is a zero mean Gaussian random variable with
variance $\sigma^2_\delta$, we have:
$\Cm = \exp(\sigma^2_\delta \Wm^2/2)$
and $(\Cm)_{kk} = \exp(-2\pi^2 k^2 \sigma^2_\delta)$, $k=-M,\dots,M$.

\paragraph{Computation of $\EE\left[\Gm_\xv\Herm\Gm_\xv\right]$}
\begin{eqnarray*}
\EE\left[\Gm_\xv\Herm\Gm_\xv\right]
&=&\EE\left[\sum_{\substack{n=0 \\ m=0}}^\infty \frac{1}{n!m!}\Deltam^n\Gm_\xvh\Herm\Wm^{\dagger\,n}\Wm^m\Gm_\xvh\Deltam^m\right] \non
&=&\sum_{\substack{n=0 \\ m=0}}^\infty \frac{1}{n!m!}\EE\left[\Deltam^n\Zm\Deltam^m\right]
\end{eqnarray*}
where $\Zm = \Gm_\xvh\Herm\Wm^{\dagger\,n}\Wm^m\Gm_\xvh$.
Now,
\begin{eqnarray*}
(\EE\left[\Deltam^n\Zm\Deltam^m\right])_{hk} 
&=& \EE\left[\delta_h^n\delta_k^m\right](\Zm)_{hk} \non
&=& \left\{
\begin{array}{ll}
\mu^{(n)}\mu^{(m)}(\Zm)_{hk} & \mbox{if $h\neq k$} \\
\mu^{(n+m)}(\Zm)_{hk} & \mbox{if $h=k$} \\
\end{array}\right.
\end{eqnarray*}
therefore
\begin{eqnarray*}
\EE\left[\Deltam^n\Zm\Deltam^m\right]
&=&\mu^{(n)}\mu^{(m)}\left(\Zm-\diag(\Zm)\right) +\mu^{(n+m)}\diag(\Zm) \non
&\hspace{-20mm}=&\hspace{-10mm}\mu^{(n)}\mu^{(m)}\Zm + \left(\mu^{(n+m)}-\mu^{(n)}\mu^{(m)}\right)\diag(\Zm)
\end{eqnarray*}
and
\begin{eqnarray}
\EE\left[\Gm_\xv\Herm\Gm_\xv\right] 
&=& \sum_{\substack{n=0 \\ m=0}} \frac{\mu^{(n)}\mu^{(m)}}{n!m!} \Zm \non
&& \hspace{-10mm}+\sum_{\substack{n=0 \\ m=0}}^\infty \frac{\mu^{(n+m)}-\mu^{(n)}\mu^{(m)}}{n!m!}\diag(\Zm)
\label{eq:EFSF}
\end{eqnarray}

The first term of (\ref{eq:EFSF}) yields
\begin{eqnarray*}
\sum_{\substack{n=0\\m=0}}^\infty\frac{\mu^{(n)}}{n!}\frac{\mu^{(m)}}{m!}\Zm
& = & \sum_{\substack{n=0 \\ m=0}}^\infty  \frac{\mu^{(n)}}{n!}\frac{\mu^{(m)}}{m!}\Gm_\xvh\Herm\Wm^{\dagger\,n}\Wm^m\Gm_\xvh \non
&\hspace{-12ex}=&\hspace{-6ex}
\Gm_\xvh\Herm \left[\sum_{n=0}^\infty \frac{\mu^{(n)}}{n!}\Wm^n\right]\Herm\left[\sum_{m=0}^\infty \frac{\mu^{(m)}}{m!}\Wm^m\right]\Gm_\xvh
\non
&\hspace{-12ex}=&\hspace{-6ex}
 \Gm_\xvh\Herm \Cm\Herm \Cm \Gm_\xvh
\end{eqnarray*}
The $(k,k)$-th element of $\Zm$ is given by
\begin{eqnarray*}
(\Zm)_{kk} &=& \sum_h (\Gm_\xvh\Herm)_{kh} (\Wm^{\dagger\,n} \Wm^m  )_{hh} (\Gm_\xvh)_{hk} \non
&=& \sum_h |(\Gm_\xvh)_{hk}|^2 (\Wm^{\dagger\,n} \Wm^m  )_{hh}\non
&=& \frac{1}{2M+1}\sum_h ( \Wm^{\dagger\,n} \Wm^m  )_{hh} \non
&=& \frac{1}{2M+1}\trace\left\{\Wm^{\dagger\,n} \Wm^m\right\}
\end{eqnarray*}
which does not depend on $k$. Thus,
$\diag(\Zm) = \frac{1}{2M+1}\trace\left\{\Wm^{\dagger\,n} \Wm^m\right\} \Id$,
and
\begin{eqnarray*}
\sum_{\substack{n=0 \\ m=0}}^\infty \frac{\mu^{(n)}}{n!}\frac{\mu^{(m)}}{m!}\diag(\Zm) && \non 
&\hspace{-40ex}=&\hspace{-20ex}
\frac{1}{2M+1}\trace\left\{\left[\sum_{n=0}^\infty \frac{\mu^{(n)}}{n!}\Wm^n\right]\Herm
\left[\sum_{m=0}^\infty\frac{\mu^{(m)}}{m!} \Wm^m\right]\right\} \Id \non 
&\hspace{-40ex}=&\hspace{-20ex}
\frac{1}{2M+1}\trace\left\{\Cm\Herm \Cm\right\} \Id
\end{eqnarray*}
Finally,
\begin{eqnarray*}
\sum_{\substack{n=0 \\ m=0}}^\infty \frac{\mu^{(n+m)}}{n!m!}\diag(\Zm)
&& \non
&\hspace{-40ex}=&\hspace{-20ex}
\frac{1}{2M+1}\trace\left\{\sum_{\substack{n=0 \\ m=0}}^\infty \frac{\mu^{(n+m)}}{n!m!}\Wm^{\dagger\,n}\Wm^m\right\} \Id \non 
&\hspace{-40ex}=&\hspace{-20ex}\frac{1}{2M+1}\trace\left\{\Ym\right\} \Id \non
\end{eqnarray*}
where
\begin{eqnarray*}
\Ym &=&\sum_{\substack{n=0 \\ m=0}}^\infty \frac{\mu^{(n+m)}}{n!m!}\Wm^{\dagger\,n}\Wm^m \non
&=&\sum_{\substack{n=0 \\ m=0}}^\infty \frac{1}{n!m!} \int_{-\infty}^{+\infty} x^{n+m} f_\delta(x) \dd x\, \Wm^{\dagger\,n}\Wm^m\non
&=&\int_{-\infty}^{+\infty}\sum_{\substack{n=0 \\ m=0}}^\infty \frac{x^n x^m}{n!m!} f_\delta(x) \Wm^{\dagger\,n}\Wm^m  \dd x \non
&=&\int_{-\infty}^{+\infty}\left[ \sum_{n=0}^\infty\frac{x^n \Wm^n}{n!}\right]\Herm\left[\sum_{m=0}^\infty \frac{x^m\Wm^m}{m!}\right] f_\delta(x)\dd x \non
&=&\int_{-\infty}^{+\infty}\exp(x\Wm)\Herm \exp(x\Wm) f_\delta(x)\dd x  \non
&=&\int_{-\infty}^{+\infty} \Id\,f_\delta(x)\dd x
= \Id
\end{eqnarray*}
Therefore,
\[ \sum_{\substack{n=0 \\ m=0}}^\infty \frac{\mu^{(n+m)}}{n!m!}\diag(\Zm) = \frac{1}{2M+1}\trace\left\{\Id\right\} \Id = \Id \]
Concluding,
\[ \EE\left[\Gm_\xv\Herm\Gm_\xv\right]  = \Gm_\xvh\Herm\Cm\Herm\Cm\Gm_\xvh + \left( 1-\frac{\trace\{\Cm\Herm\Cm\}}{2M+1}\right)\Id \]
and, if $\Cm$ is real,
\[ \EE\left[\Gm_\xv\Herm\Gm_\xv\right]  = \Gm_\xvh\Herm \Cm^2 \Gm_\xvh + \left(1-\frac{\trace\{\Cm^2\}}{2M+1}\right)\Id \]


\end{document}